\documentclass[useAMS,usenatbib,onecolumn]{mn2e}
\usepackage{epsfig}
\usepackage{graphicx}
\usepackage{amsmath}  
\newcommand{\etal}{{et~al.}~}
\newcommand{\lsim}{\,\lower2truept\hbox{${<\atop\hbox{\raise4truept\hbox{$\sim$}}}$}\,}
\newcommand{\gsim}{\,\lower2truept\hbox{${>\atop\hbox{\raise4truept\hbox{$\sim$}}}$}\,}
\newcommand{\beq}{\begin{equation}}
\newcommand{\eeq}{\end{equation}}

\newcommand{\WMAP}{$WMAP$}
\newcommand{\fastica}{{\sc{fastica}}}

\newcommand{\vect}[1]{{\mathbfit{#1}}}

\def\aas{{\sl Astron. \& Astrophys.\ Suppl.\ }}
\def\apj{{\sl Astrophys.\ J.\ }}
\def\apjl{{\sl Astrophys.\ J.\ Lett.\ }}
\def\apjs{{\sl Astrophys.\ J.\ Supp.\ }}

\def\ieeespl{{\sl IEEE\ Signal\ Processing\ Lett.\ }}

\def\mnras{{\sl MNRAS\ }}

\def\nn{{\sl Neural Networks\ }}

\title[Foreground analysis of the \WMAP~3yr data with \fastica]
{Foreground analysis of the \WMAP\ three-year data with \fastica}

\author[Bottino \etal]{M. Bottino$^{1,2}\footnote{E-mail:bottino@mpa-garching.mpg.de}$,
A.~J. Banday$^2$, D. Maino$^1$\\
$^{1}$ Dipartimento di Fisica, Universit\`a di Milano, Via Celoria 16, I-20133, Milano, Italy\\
$^{2}$ Max-Planck Institute f{\"u}r Astrophysik, Karl-Schwarzschild Str. 1, D-85748, Garching, Germany\\
}

\pagerange{\pageref{firstpage}--\pageref{lastpage}}
\pubyear{2006}

\begin{document}

\maketitle

\label{firstpage}

\begin{abstract}

We present an analysis of the foreground emission present in the
\WMAP\ 3-year data as determined by the method of Independent
Component Analysis. The \WMAP\ data averaged at each frequency are
used together with the standard foreground emission templates as
inputs to the \fastica\ algorithm. The returned coefficients can be
interpreted as coupling coefficients between the \WMAP\ data and
foreground templates.  These results are then used to infer the
spectral behaviour for three foreground components -- synchrotron,
anomalous dust-correlated emission and free-free. For the first two
components, we find values consistent with previous results although
slightly steeper. We confirm the inconsistency in the scaling between
the H$\alpha$ template and free-free emission at K- and Ka-bands where
an electron temperature of $\sim 4000$~K is indicated. We also see
evidence of significantly flatter spectral behaviour to higher
frequencies than expected theoretically and previously noted by
\citet{dobler_etal_2008a}, but only when analysing the $Kp2$ sky
coverage.

We further apply \fastica\ \lq iteratively', using data pre-cleaned
using foreground templates scaled to the \WMAP\ frequencies by
coupling coefficients determined by a prior \fastica\ analysis.  This
multi-frequency analysis allows us to determine the presence of
residual foreground emission not traced by the templates. We confirm
the existence of a component spatially distributed along the Galactic
plane and particularly enhanced near the center (the \lq \WMAP\
haze'). This emission is less extended when using the \WMAP\ K-Ka data
as the synchrotron template confirming that it can be considered a
better template for foreground cleaning of the \WMAP\ data. However
its use complicates the physical interpretation of the nature of the
foreground emission and residuals since it contains a mixture of
several, physically distinct emission mechanisms.

The good agreement between the extracted CMB component and previous
results, as well as the low amplitude of residual foreground emission
make \fastica\ a viable tool to infer foreground emission properties,
via template fitting, and the CMB amplitude.

\end{abstract}

\footnotetext{E-mail: bottino@mpa-garching.mpg.de}

\section{Introduction}
\label{intro}

Measurements of the Cosmic Microwave Background (CMB) provide
cosmological information by probing the early universe
some 380,000 years after the Big Bang.
However their precision is limited by the
contamination of the cosmological signal due to emission from
local astrophysical sources such as our own Galaxy.
In particular the Galactic contamination arises from
three well-understood processes: synchrotron emission, 
free-free emission (or \emph{thermal bremsstrahlung}) 
and thermal dust emission. In addition, there appears
to be an enhanced dust correlated emission at microwave wavelengths
that is inconsistent with the thermal mechanism, and the exact nature
of which is still open to debate.
Theoretically, these components can be
distinguished on the basis of their own spectral properties and spatial
distribution. The best window for observing the temperature anisotropies lies close
to 70 GHz, where the integrated diffuse foreground emission seems to
display a minimum. This
has permitted accurate mapping of the CMB intensity pattern, without
the need to excise too much sky coverage from the analysis due to these
foregrounds (except in regions near to the Galactic plane and
associated with bright point sources).
However, to reach the accuracy required for precision cosmology ($\sim
\mu$K), component separation
techniques and/or foreground cleaning methods are needed. 

A component separation technique is one that attempts to decompose
multi-frequency input data into separate physical emission components,
one of which, for our purposes, must be the CMB. However, the
technique may not be successful particularly in regions of the sky
where the foreground spectral dependencies are complex, or where the
physical emissions have complex morphology and are well-mixed. A
foreground cleaning method in the context of this paper simply
attempts to generate a map of the CMB as clean of foreground emission
as possible, with no concern for the detailed separation of the
foreground emission into physical components. It should be clear that
a component separation algorithm may have limited success, providing
an adequate reconstruction of the CMB but not of the foreground
emission, then effectively consituting a foreground cleaning method.

In this paper we will consider the detailed application of one technique --
\emph{Independent Component Analysis} (ICA) -- to data from the
\WMAP\footnote{\emph{Wilkinson Microwave Anisotropy Probe}} satellite,
focusing our attention on the properties of diffuse Galactic
foregrounds at
high and intermediate latitude as traced by specific templates
of the emission.
Of course, \fastica\ is only one of a number of 
methods that have been developed for the application to data at
microwave wavelengths in order to (at least) extract a map of the CMB
sky with minimal foreground residuals.
An inexhaustive list of other techniques applied to the \WMAP\ data includes
linear combination methods \citep{bennett_etal_2003,tegmark_etal_2003,park_etal_2007},
WIFIT \citep{hansen_etal_2006},
FGFit \citep{eriksen_etal_2006},
SMICA \citep{patanchon_etal_2005} and
CCA \citep{bonaldi_etal_2007}.
The \WMAP\ team has also made use of the Maximum Entropy Method (MEM)
\citep{bennett_etal_2003,hinshaw_etal_2007}
to study the detailed properties of the Galactic foregrounds.
However, since MEM is not able to produce a CMB
map that is easy to use for cosmological analyses, the \WMAP\ data was cleaned
from foregrounds by applying a template fitting
algorithm --  
essentially a cross-correlation method also used by
\citet{davies_etal_2006} to study regional variations in the
foreground emission.

A general summary of the paper is as follows.
We first provide a simple introduction to ICA and describe
the use of templates of foreground emission in such an analysis,
making a comparison to more standard template fitting techniques.
We then describe the data used in this paper for the foreground
analysis. Then we calibrate the performance of the method using Monte Carlo
simulations and by comparison to the simpler template based method.
Sections~\ref{haslam_analysis} and \ref{K-Ka_analysis} 
then attempt to improve our knowledge of the spectral and physical
properties of the foreground components based on the scaling
properties of the templates. 
An assessment of the efficiency of
the method is then afforded by studying the power spectrum of the cleaned 
CMB data, whilst a final iterative application of ICA attempts
to improve the removal of foreground contamination from the data.
Section~\ref{discussion} summarises our main conclusions.

\section{\fastica\ and its use for foreground component studies}
\label{ica_templates}

ICA is a \emph{blind} component separation method: it 
works without any knowledge about the spectral and spatial properties
of the components to be
recovered. A particular application of this method is {\fastica}
as originally proposed by \citet{hyvarinen_1999,hyvarinen_oja_2000} 
and applied to the analysis of CMB total intensity data in
\citet{maino_etal_2002} 
and more recently to polarization in \citet{maino_etal_2002} and
\citet{baccigalupi_etal_2004}. 

The component separation problem is formalized assuming that the sky
radiation at a given frequency $\nu$
can be described as a linear combination of $N$ different physical
processes whose frequency and spatial dependencies can
be factorized into two terms:
\beq
\tilde{x}(\vect{r},\nu) = \sum_{j=1}^N \bar{s}_{j}(\vect{r})f_{j}(\nu)\, .
\eeq
Generally the radiation is observed by a combination of an optical system
and a M-channel measuring instrument. In what follows,
we will assume that the observations can then be described
as the convolution of the sky signal with a
frequency dependent azimuthally symmetric beam response and 
a bandpass $t_{\nu}(\nu')$ (the frequency response of the
channel) which is effectively a delta function. 
The former assumption will generally be violated, but
for \WMAP\ the effect of multiple observations of a given
point on the sky with the detectors in different relative orientations
(due to the satellite scan pattern)
results in an effectively symmetric beam.
This may not be the case for the {\em Planck} satellite. The latter assumption
can also be improved on by defining an effective frequency
for a given detector depending on the spectral behaviour of the
impinging radiation. Nevertheless, with our simplifying
assumptions we can
define the scaling coefficients 
$a_{\nu j}=f_{j}(\nu)$, and construct the
relevant $M\times N$ mixing matrix {\bf A}. 
If we introduce the
instrumental noise {\bf $\epsilon$}, and we assume that the beam function is 
frequency independent $B(\vect{r},\nu)=B(\vect{r})$,
the observed signal for each position $r$ is expressed by 
\begin{equation}
{\bf{x}}(\vect{r}) = {\bf{A}} {\bar{\bf{s}}}(\vect{r}) * B(\vect{r}) +
{\bmath{\epsilon}}(\vect{r}) = {\bf{A}} {\bf{s}}(\vect{r}) + {\bmath{\epsilon}}(\vect{r})\, ,
\end{equation}
where {\bf x} and  {\bf $\epsilon$}
are vectors with $M$ rows, and the star represents the convolution 
of the beam PSF with the sky signals $\bar{\bf s}$, indicated simply as 
${\bf s}$ afterward.

The ICA approach then estimates both the
matrix {\bf A} and the vector {\bf s} from the data {\bf x}
by assuming that
\begin{itemize}
\item the sources {\bf s} are random independent processes on the map
\item the components of the vector {\bf s} with at most one exception have
      non-Gaussian distributions.
\end{itemize}
The problem is solved by
finding a linear transformation {\bf W} so that the components of the transformed vector
${\mathbf{y}}={\mathbf{W}}{\mathbf{x}}$
are independent. Using the \emph{Central Limit Theorem} the
independence is achieved maximizing the non-Gaussianity of the
variables. This corresponds to the maxima of their 
\emph{neg-entropy} which in the \fastica\ implementation is generally approximated by three non-linear 
functions, i.e. $p(u) = u^3$, $t(u) = 
{\rm tanh}(u)$ and $g(u) = u {\rm exp}(-u^2)$ where $u$ are the
principal component projected data \citep{hyvarinen_1999,hyvarinen_oja_2000}.
These functions depend on the statistics of the 
independent signals which are assumed to be unknown.
Generally, $p$, 
which corresponds to the kurtosis, should be used for sub-Gaussian 
components but it is strongly sensitive to outliers in the distributions; 
$g$ is for super-Gaussian signals while $t$ is a general purpose function 
\citep{hyvarinen_1999}.
Although {\fastica} is a non linear algorithm, the returned
components are defined as linear combinations of the input data,
whose weights are the elements of the matrix ${\mathbf{W}}$:
\begin{equation}
y_{j} = \sum_{i=1}^{M} w_{ij} x_{\nu_{i}}. \\
\end{equation}
\label{y_comp}
Once the matrix {\bf W} is determined, the independent components are
recovered via the relation:
\begin{equation}
x_{\nu_{i}} = \sum_{j=1}^{N} w_{ij}^{-1} y_{j} \\
\end{equation}
with $i=1,...N$.
 
\fastica\ has been utilised in component separation studies for both
the \emph{COBE}-DMR and \WMAP\ multi-frequency observations of the
microwave sky in \citet{maino_etal_2002} and \citet{maino_etal_2006}
respectively. However, in general it has been found that the 
foreground components returned by the method are not optimal -- a
fully reliable component separation is not achieved. Nevertheless, one
of the output maps of the analysis can, in general, be associated explicitly and
robustly with the CMB signal.  In an alternative approach, \citet{maino_etal_2002}
proposed the analysis of a single microwave sky map in combination
with observations at wavelengths where only single foreground physical
emission processes dominate.  These various foreground templates are
considered to be representative models for the morphology of the
foreground emission at that wavelength (see section \ref{data} for
details of the templates used in this analysis), except for a
scale-factor to be determined.
Application of the {\fastica} algorithm then provides
information on foreground emission, not directly from the spectral
properties of the reconstructed and ambiguous Galactic components, but
from the recovered CMB component.  Referring to Eq. \ref{y_comp}, the
reconstructed CMB component is defined as a weighted linear
combination of the input data, namely the microwave sky map and
and putative templates of Galactic emission. Those coefficients associated with the
foreground templates define the effective contamination of the CMB
anisotropy signal by each foreground.  
After normalising the weights so that the factor associated with the microwave sky map is
unity, we can then interpret the modified template
weights as correlation coefficients between the foregrounds
and a given channel of the microwave data.  Therefore with
\fastica-derived CMB sky maps at different frequencies, we can derive
the frequency dependence of the coupling coefficients associated with
a given template or physical foreground component. 

However, for such an analysis it is necessary to focus attention on 
the high and intermediate latitudes for two reasons.  Firstly, it is
likely that the physical conditions within the Galactic plane are
different to those at higher latitudes, so that the assumption that
the templates can be scaled to a given wavelength by a single factor
(or spectral index) is almost certainly invalidated, thus a global
\fastica\ analysis will be compromised.  Secondly, the complex
morphology of the Galactic plane results in non-negligible
correlations between the different emission mechanisms, thus violating
one of the assumptions of the ICA approach.
 
Such an analysis is then directly analogous to the simple template
fitting scheme commonly used in the field. Therefore, we also utilise
the technique here as a convenient point of reference for our
\fastica\ results.  In general, the cross--correlation measure,
$\alpha$, between a data vector ${\bf d}$ and a template vector ${\bf
t}$ can be measured by minimising:
\begin{equation}
\chi^2 = ({\bf d}-\alpha {\bf t})^T \cdot {\bf M}^{-1}_{SN}\cdot ({\bf d}-\alpha {\bf t}) = {\bf \tilde{d}}^T \cdot {\bf M}^{-1}_{SN}\cdot {\bf \tilde{d}}
\end{equation}
where ${\bf M}_{SN}$ is the covariance matrix including both signal
and noise for the template--corrected data vector ${\bf \tilde{d}}
\equiv {\bf d} - \alpha {\bf t}$. 
In the case of $N$ different foreground components, we have
the simple system of linear equations ${\bf Ax}={\bf b}$, where
\begin{gather}
A_{kj}=\mathbf{t}^T_k \cdot \mathbf{M}_{\textrm{SN}}^{-1} \cdot
\mathbf{t}_j, \notag \\ 
b_k = \mathbf{t}^T_k \cdot \mathbf{M}_{\textrm{SN}}^{-1} \cdot \mathbf{d}, \notag \\ 
x_k = \alpha_k.
\end{gather} 
The correlation method can also be extended to include various
constraints on the data, eg. fixed dust or free-free spectral
indices. This is the case described by \citet{hinshaw_etal_2007} in
the analysis of the 3 year data of \WMAP\, but we impose no such
constraints here.

Once the scaling factors are computed they can be used to clean the
\WMAP\ data from the estimated Galactic contaminations. The residuals
then provide another characterization of the accuracy of the cleaning
process.  Indeed, residual foreground components are expected that are
not well correlated with the adopted foreground templates.  They are
most easily interpretated as a consequence of the main limitation of
{\fastica} as currently applied, namely the de-facto assumption of a
fixed spectral index for each physical foreground component over the
analysed sky coverage.  We will show that an \lq iterative' application
of the \fastica\ algorithm to multi-frequency pre-cleaned microwave
data can provide additional insight into such residuals and an
additional estimate of the CMB component.  The outcome of such an
iterative stage will include an improved CMB sky map, plus component
maps that represent foreground residuals, although it is unlikely that
they can be unambiguously assigned to specific physical components.

\section{DATA USED IN THE ANALYSIS}
\label{data}
The current analysis seeks to 
clean the \WMAP\ data from the main Galactic foreground
emissions by applying the {\fastica} algorithm 
to the observed \WMAP\ sky signal at each frequency together with
appropriate templates for each component.
The current implementation of the {\fastica} 
algorithm assumes that each input sky map has the same spatial
resolution. 
Thus we perform our analysis on sky maps convolved from their original resolution to 
an effective $1^{\circ}$ Gaussian beam.
Since regions close to the Galactic plane are the most seriously contaminated by
foregrounds and the spectral and spatial nature of the integrated emission is
complex, we exclude them according to the $Kp2$ and $Kp0$ masks
provided by the \WMAP\ team \citep{bennett_etal_2003}. 
The Galactic part of the masks as well as 
the point-source exclusion regions are considered
sufficient and no modifications are made.

\subsection{\WMAP\ data}
\label{wmapdata}
The \WMAP\ satellite \citep{hinshaw_etal_2007} observes the sky with
ten so-called differencing assemblies (DAs), with frequency dependent resolution of
approximately $0.23^{\circ}$ to $ 0.93^{\circ}$. 
The frequency ranges  from $\sim$23~GHz (K-band) up to
$\sim$94~GHz (W-band): there are two channels in the Q-
and V-bands, and four channels in the W-band
whilst there is only one channel at each of the K- and Ka-bands.
The corresponding sky map data are available on the LAMBDA
website\footnote{\emph{ Legacy Archive for Microwave Background Data
Analysis} -- http://lambda.gsfc.nasa.gov/.} 
in a  HEALPix\footnote{http://healpix.jpl.nasa.gov.} 
pixelisation scheme, with a pixel resolution parameter of $N_{side}=512$.

We have smoothed each of the sky maps to an effective resolution of
$1^{\circ}$ (initially deconvolving the azimuthally symmetric 
beam profile for each DA), 
then we have combined the maps in
the Q-, V- and W-bands using a simple average over the corresponding
DAs in order to generate a single map for each frequency band.
The effective central frequency of each DA depends on the spectrum of
the emission being considered. For the band-averaged data,
we simply adopt the values 23, 33, 41, 61 and 94 GHz for K- through
W-band. 
Finally, we converted the \WMAP\ data from thermodynamic temperature to
brightness (antenna) temperature
units, in order to make the data consistent with the
templates. 

\subsection{Synchrotron templates}
\label{sync}

Synchrotron emission arises from the acceleration of cosmic ray electrons in the
magnetic field of our Galaxy.
This is the dominant Galactic emission at frequencies lower than $10~$GHz,
with an intensity that depends on the
Galactic magnetic field, on the energy spectrum of the electrons and on
their spatial distribution. The resulting 
synchrotron emission is generally described by a power-law 
spectrum
\begin{equation}
T(\nu) \sim \nu ^{-(p+3)/2} = \nu^{-\beta_s}
\end{equation}
where the index $p$ is the spectral index of the energy spectrum of
the electrons, and the parameter $\beta_s$ is the synchrotron spectral
index.
The spectral behaviour changes with the frequency as a consequence
of the details of the cosmic ray electron propagation, including 
energy loss, and degree of confinement.

The 408 MHz radio continuum all-sky map of \citet{haslam_etal_1982} is
dominated by synchrotron emission away from the Galactic plane. 
Some contribution from free-free emission is observed at the 
lowest latitudes, for example \citet{reich_reich_1988}
estimate the relative fraction at $b=0^{\circ}$ to be of order 10--20\%.
However, the contribution at intermediate and high latitudes is
effectively negligible and hence the 408 MHz sky map is generally assumed as
the default template for the synchrotron foreground component. 
Studies indicate a range of spectral indices 
$\beta_s$ from 2.3 to 3.0 between 408 MHz and 1420
MHz \citep{reich_reich_1988} and the presence of spurious baseline
effects in the survey \citep{davies_etal_1996} that affect the
spectral index determination for weaker features.
The combination of these effects, plus the large frequency gap 
between 408 MHz and microwave wavelengths, 
does suggest that the sky map scaled by a single
effective index may not be representative
of the synchrotron emission at \WMAP\ frequencies.
Nevertheless, its use for foreground studies is well established
in the CMB literature and we retain it here.
However, we also use the 
difference between the K and Ka \WMAP\ data as a synchrotron template, 
as suggested by \citet{hinshaw_etal_2007}.
This at least in part compensates for possible errors
introduced by using the Haslam template,
presumably accounting for the change in morphology of the emission
at microwave frequencies. Moreover, using this template with a
single fitted scaling per frequency is likely to be sufficiently accurate
even given modest departures from a single spectral index. 
It is also likely to be a good choice
because the intrinsic systematic measurement errors are smaller
than for the Haslam map.
However, the interpretation of the scalings is complicated by the fact
that the difference map must also include a free-free component,
and probably a contribution from the anomalous dust correlated emission.

\subsection{H$\alpha$ free-free template}
\label{freefree}

The free-free emission arises from electron-ion scattering. Its
spectrum is
generally described by a power law $T_A\sim \nu^{-\beta_{ff}}$ 
with $\beta_{ff}=2.15$ at high-frequency ($\nu > 10$ GHz) and $\beta_{ff}=2$
at low frequency, due to optically thick self-absorption. 
Because the free-free emission
does not dominate the sky intensity at any radio frequency, our knowledge about
it come from H$\alpha$ radiation: both are connected to the ionized interstellar
medium and proportional to the \emph{Emission
Measure} $EM= \int n_e^2 dl$, where $n_e$ is the electron volume density.

\citet{finkbeiner_2003} has produced an all-sky
H$\alpha$-map by assembling data from several surveys: the Wisconsin
H$-$Alpha Mapper (WHAM) \citep{haffner_etal_2003}, the Virginia Tech Spectral-Line Survey (VTSS)
\citep{dennison_etal_1998}, and the Southern H-Alpha Sky Survey Atlas
(SHASSA) \citep{gaustad_etal_2001}. The final resolution of the map is
asserted to be equal to $6$ arcmin. 
At intermediate and lower Galactic latitudes the absorption of the
H$\alpha$ emission by dust is significant and this requires correction
if the template is to be used as a tracer of the free-free emission.
\citet{dickinson_etal_2003} have used the
\citet{schlegel_etal_1998} (SFD) dust map derived from the IRAS
100 $\mu m$ all-sky survey corrected to a fixed temperature of
$18.3~K$ to estimate the absorption correction. Its amplitude in
magnitudes at the H$\alpha$ wavelength is equal to $A({\rm H}\alpha) = (0.0462 \pm
0.0035)D^{T}f_{d}$, where $D^{T}$ is the SFD temperature-corrected
100$~\mu$m intensity in MJy sr$^{-1}$ and $f_{d}$ is the fraction of
dust in front of the H$\alpha$ in the line of sight. A value of $f_{d}
\sim 0.5$ is expected under the assumption that the  ionised gas and dust
are uniformly mixed along the line of sight. However,
\citet{dickinson_etal_2003} have presented evidence for a lower value
$f_{d} \sim 0.3$, whilst \citet{banday_etal_2003}
suggest that zero
correction is required for high Galactic latitudes ($|b| >
20^{\circ}$). 
This is the value that we have adopted in our analysis.

After correction the brightness temperature $T_{b}$  
can be related to the EM (in units of cm$^{-6}$~pc)
using $T_{b} \propto T_{e}^{-0.35} \nu^{-2.1} \times EM$.
$T_{e}$ of ionised
gas varies as $\nu^{\sim 0.7}$ in the conversion of H$\alpha$
intensity to brightness temperature at microwave frequencies.  For the
\emph{WMAP} bands K, Ka, Q, V and W this corresponds to 11.4, 5.2,
3.3, 1.4 and 0.6~$\mu$K~R$^{-1}$ respectively for $T_{e}=8000$~K; see
\citet{dickinson_etal_2003} for details.

\subsection{Dust templates}
\label{dust}

Thermal dust emission is the dominant foreground component at frequencies 
in the range $\sim$\,100 -- 1000~GHz. 
Its emissivity is generally approximated by a modified black body law
\begin{equation}
I(\nu) \sim \nu^{\alpha_d}B_{\nu}(T_d)
\end{equation}
where $\alpha_d$ is the emissivity index and $B_{\nu}(T_d)$ the
blackbody emissivity at a dust temperature $T_d$.
In fact, different type of dust grains (silicates and carbonates)
exhibit different spectra characterized by a range of emissivity
indices  and dust temperatures.
\citet{finkbeiner_etal_1999}
developed a series of models (FDS) based on the \emph{COBE}-DIRBE 100 and 240 $\mu$m  maps
tied to the \emph{COBE}-FIRAS spectral data in the range 0.14 to
3.0~mm.
Model 8 is considered to be the preferred model at microwave wavelengths and 
comprises two components with spectral indices
$\alpha_{d1} = 1.67$ and $\alpha_{d2} = 2.7$ and temperatures 9.4\, K
and 16.2\, K respectively. 
Over the \WMAP\ range of frequencies, the thermal dust emission
in antenna temperature is sometimes considered to scale 
with a power-law dependence $T_A\sim \nu^{\beta_{d}}$, with values
of 1.7 -- 2.2 for ${\beta_{d}}$ assumed by different authors.
We adopt the FDS8 predicted emission at 94
GHz as the reference template for dust emission and compute a single
global scaling between this template and the \WMAP\ frequencies.
The map has a claimed nominal resolution of $6.1$ arcmin.

At frequencies between $\sim$\, 10 and 100 GHz considerable evidence
has been accrued for a foreground component with a synchrotron-like spectrum
that is at least partially spatially correlated with the thermal dust 
emission
\citep{kogut_etal_1996a, kogut_etal_1996b, banday_etal_2003, maino_etal_2003,
leitch_etal_1997, deoliveira-costa_1998a, bennett_etal_2003, 
hinshaw_etal_2007, finkbeiner_etal_2002, casassus_etal_2004}.
Several  models have been advanced  to explain the origin of the
so-called anomalous dust emission, although a definitive model 
remains unclear. 
The \WMAP\ science team prefer an
interpretation in which the emission is flat-spectrum synchrotron
emission that originates in star-forming regions close to the plane.
The more commonly accepted theoretical model is that the emission
arises from spinning dust grains, ie. the rotational modes of
excitation of very small dust grains with a high rotation velocity 
\citep{draine_etal_1998a, draine_etal_1998b}. 
We clearly need an accurate template for this anomalous emission
to model the foreground amplitudes at \WMAP\ frequencies, but this
remains elusive. 
\citet{finkbeiner_2004} and \citet{davies_etal_2006} 
have proposed that the FDS template modulated by some power-law
of the dust temperature provides a better fit than the unperturbed
sky map. However, \citet{bonaldi_etal_2007} suggest that although 
the anomalous emission is tightly correlated with thermal dust, 
the correlation is not perfect. Nevertheless, we expect that the FDS
model will provide a representative template for our analysis.

\section{MONTE CARLO SIMULATIONS AND CALIBRATION OF METHOD}
\label{simulations_haslam}

As the first step of our analysis, we have tested {\fastica} 
using realistic simulations of the \WMAP\ observations
to calibrate the accuracy of the method for template fitting.
Specifically, we performed 1000 simulations of the microwave sky
at each of the 5 \WMAP\ frequencies, each containing a realisation of 
the CMB signal, the Galactic foreground emission, and instrumental
noise appropriate to the specific channel. The study was undertaken
at an effective resolution of $1^{\circ}$.

Each simulation was used as an 
input to {\fastica}, together with the foreground emission templates,
and the scaling factors computed.
Appendix~\ref{appendix} provides full details of their statistical
distribution as a function of both the applied Galactic cut ($Kp2$ or $Kp0$),
and the \fastica\ non-linear function.
In summary, the main conclusions are as follows:
\begin{itemize}
  \item the $t$-function is inappropriate for template fitting --
     the statistical distributions of the returned coupling
     coefficients are asymmetric and highly biased with respect to
     the input values 
  \item the $p$- and $g$- functions provide similar results --
    the distributions of the coupling coefficients are
    well-described by Gaussians with a weak bias of the mean
    that is essentially insignificant for our studies. 
  \item there is evidence of cross-talk, specifically an
     anti-correlation, between the synchrotron and dust coefficients
     when either the Haslam or K-Ka templates are used to characterise
     the synchrotron emission. In the latter case, correlation between
     the template and free-free emission is also seen.
  \item the results from the simple $\chi^2$ analysis indicate that the
     uncertainties are significantly larger than for the \fastica\ method,
     substantially so when compared to the $p$-function. Moreover, the
     method seems to demonstrate more cross-talk between components
     than seen for the \fastica\ analysis.
\end{itemize}

The uncertainties for the scaling factors presented in later
sections of the paper are derived directly from these simulations.

\section{ANALYSIS WITH THE HASLAM MAP AS SYNCHROTRON TEMPLATE.}
\label{haslam_analysis}

For our initial analysis we consider template fits to the \WMAP\ data
for all five frequency bands using {\fastica} and the Haslam map as
the model for the synchrotron emission.
Table~\ref{table_1deg_coeffs_haslam} summarises the results determined
using both $p$- and $g$-functions for the $Kp2$ and $Kp0$ sky
coverages. The corresponding $Kp2$ results based on a simple
$\chi^2$ analysis (see Appendix~\ref{appendix} for details)
are also shown, together
with the template fit coefficients used by \WMAP\ for their first year
foreground correction \citep{bennett_etal_2003}. We can make the
following general observations based on these results.

\begin{table} 
\begin{center} 
\begin{tabular}{l cccccc } 
\hline
\hline 
& \multicolumn{2}{c}{\bf{synchrotron}} &  \multicolumn{2}{c}{\bf{dust}} &  \multicolumn{2}{c}{\bf{free-free}}  \\ 
\hline
&$Kp2$&$Kp0$& $Kp2$&$Kp0$&$Kp2$&$Kp0$\\
\hline
&\multicolumn{6}{c}{\bf{{\fastica} - function p }}\\ 
\hline 
$K$   & $6.64\pm0.44$&$5.46\pm0.62$ &$5.77\pm0.26$& $4.94\pm0.33$ & $8.58\pm0.39$ &$6.70\pm0.63$\\ 
$Ka$  & $2.00\pm0.43$&$1.68\pm0.61$ &$2.02\pm0.26$& $1.28\pm0.32$ & $4.31\pm0.38$ &$2.68\pm0.62$\\
$Q$   & $1.01\pm0.42$&$0.87\pm0.60$ &$1.08\pm0.26$& $0.42\pm0.31$ & $2.89\pm0.38$ &$1.36\pm0.61$\\ 
$V$   & $0.26\pm0.40$&$0.26\pm0.55$ &$0.60\pm0.24$& $0.03\pm0.28$ & $1.39\pm0.36$ &$0.01\pm0.57$\\ 
$W$   & $0.05\pm0.35$&$0.05\pm0.49$ &$0.95\pm0.21$& $0.46\pm0.25$ & $0.67\pm0.32$ &$-0.49\pm0.52$\\ 
\hline
&\multicolumn{6}{c}{\bf{{\fastica} - function g }}\\ 
\hline 
$K$   & $6.52\pm0.54$&$5.66\pm0.80$ &$6.07\pm0.29$& $5.71\pm0.41$ & $8.52\pm0.54$ &$6.60\pm0.82$\\ 
$Ka$  & $2.05\pm0.53$&$1.87\pm0.79$ &$2.09\pm0.28$& $1.66\pm0.40$ & $4.21\pm0.53$ &$2.75\pm0.80$\\
$Q$   & $1.09\pm0.52$&$1.08\pm0.78$ &$1.11\pm0.28$& $0.69\pm0.40$ & $2.79\pm0.53$ &$1.47\pm0.79$\\ 
$V$   & $0.39\pm0.49$&$0.56\pm0.72$ &$0.58\pm0.27$& $0.19\pm0.36$ & $1.31\pm0.50$ &$0.19\pm0.76$\\ 
$W$   & $0.18\pm0.43$&$0.31\pm0.65$ &$0.93\pm0.23$& $0.59\pm0.32$ & $0.61\pm0.43$ &$-0.36\pm0.67$\\ 
\hline
\hline 
&\multicolumn{6}{c}{\bf{$\chi^2$ analysis}} \\
\hline
$K$   & $5.96\pm0.57$&&$6.38\pm0.29$&&$8.00\pm0.62$&\\
$Ka$  & $1.83\pm0.56$&&$2.25\pm0.29$&&$3.71\pm0.60$&\\
$Q$   & $0.94\pm0.54$&&$1.24\pm0.28$&&$2.32\pm0.58$&\\
$V$   & $0.18\pm0.49$&&$0.69\pm0.25$&&$0.88\pm0.53$&\\
$W$   & $0.01\pm0.37$&&$1.02\pm0.19$&&$0.23\pm0.41$&\\
\hline 
&\multicolumn{6}{c}{\bf{Bennett et al. values (constrained: $\beta_s = 2.7$; $\beta_{ff} = 2.15$)}} \\
\hline
$K$  &$-$ &&$  -  $&&$  - $&\\
$Ka$ &$-$ &&$  -  $&&$  - $&\\
$Q$  &$1.01$ &&$  1.04  $&&$ (1.92)  $&\\
$V $ &$0.34$ &&$  0.62  $&&$ (0.82)  $&\\
$W $ &$0.11$ &&$  0.87  $&&$ (0.32)  $&\\
\hline
\hline
\end{tabular}
\caption{\emph{Values of the coupling coefficients in antenna
temperature units determined
between the 3-year \WMAP\ data and three foreground emission templates, 
at $1^{\circ}$ resolution. The Haslam 408~MHz map is
adopted as the synchrotron template. The {\fastica} analysis
is performed using the two non-linear functions $p$ and $g$ 
and for the $Kp2$ and $Kp0$ masks.
The corresponding $Kp2$ results for a simple $\chi^2$ analysis
are provided for comparison. In addition, we provide
the values from the \citet{bennett_etal_2003} fits
to the Q,V and W-bands performed with constraints imposed
on the synchrotron and free-free spectral indices. These
are the coefficients adopted by \WMAP\ for their first-year foreground
corrections. 
The units are $\mu$K/K for synchrotron, mK/mK
for dust and $\mu$K/R for free-free emission respectively.}}
\label{table_1deg_coeffs_haslam}
\end{center}
\end{table}

The frequency dependence of the coefficients 
follows the generally expected trends, and in particular, the 
synchrotron and free-free coefficients decrease with increasing
frequency. 
The situation with dust emission is more complex:
the dust emission decreases from $94~\rm{GHz}$ to $61~\rm{GHz}$
as would be expected for thermal dust emission.
However, there is a strongly increasing contribution from a dust
correlated component with decreasing frequency below $61~\rm{GHz}$.
This presumably constitutes further evidence 
for a new emission mechanism for dust, the precise physical nature of
which remains unknown.
Overall, however, there is 
a minimum contribution from the diffuse Galactic foregrounds at
$\sim 61~\rm{GHz}$, which can be regarded as defining the optimal range
for CMB temperature measurements.

The amplitude of the derived coefficients depends on the extent 
of the mask applied to the data.
For all the foreground components, 
almost without exception, the coefficients derived with the $Kp0$ sky
coverage are less than for $Kp2$. This systematic trend presumably
results from genuine spectral variations of the foregrounds on the
sky. Indeed, this behaviour is particularly  
noticeable for the free-free emission, and may reflect the fact that 
the $Kp0$ mask excludes strong emission regions
near the Galactic plane, 
whilst the free-free emission remains weak at higher latitudes.
These strong emission regions may also  manifest
real variations in scaling dependence relative to the H$\alpha$
template due to variations in the dust absorption or in the temperature of the ionised gas
in the medium latitude region present in the $Kp2$ mask but removed by
$Kp0$. 

The {\fastica} numbers are also in good statistical agreement with our
own simple $\chi2$ results, provided only for the $Kp2$ sky coverage.
Interestingly, the synchrotron and dust amplitudes are systematically
lower (higher) for the $\chi2$ method. However, these results in
general reflect the weak cross-talk seen between the fitted amplitudes
in Appendix~\ref{appendix}. The free-free amplitudes are also
typically 1-$\sigma$ higher for the {\fastica} analysis.

Similarly, the constrained fits from \citet{bennett_etal_2003} are of
comparable amplitude for the synchrotron and dust coefficients. These
values were derived by fixing the synchrotron spectral index at a
value of 2.7, and the free-free to 2.15 for frequencies above
Q-band. The former constraint
explains why the synchrotron amplitude remains higher at V- and W-band
for the first-year \WMAP\ corrections, although the Q-band amplitude is
in excellent agreement with the {\fastica} numbers. It is curious to
note that the constrained synchrotron index adopted is inconsistent with the
Q-band amplitude, from which one can infer a spectral index of $\sim
3$ between 408~MHz and 61~GHz. However, \citet{bennett_etal_2003}
specifically state that their template method is not particularly
physical, but removes foregrounds outside the $Kp2$ cut at the
required level. 

Finally, we would like to comment on the impact of dipole subtraction
for the analysis. Specifically, the results we have presented here,
including those of \citet{bennett_etal_2003}, do not subtract a
best-fit dipole amplitude from the data and templates before fitting
the template amplitudes. Since other independent analyses,
eg. \citet{davies_etal_2006}, do prefer to remove the dipole, we have
tested its impact on the results.  In fact, we find that the dust and
free-free coefficients remain largely unchanged, whilst the synchrotron
values become negative at high frequencies. It may be that this
is connected with the projection of the North Polar Spur in the Haslam
template onto the best-fit dipole computed for that template.

\subsection{Spectral index of foreground emissions.}
\label{haslam_spectra}

We utilise the derived frequency dependence of the scaling factors to
parameterise the spectral behavior of the foreground emission
components.  For each component, we fit the corresponding coefficients
with a power law model of the form
$A_{norm}(\nu/\nu_{0})^{-\beta}$. $A_{norm}$ is the amplitude of the
emission of a specific physical component at the reference frequency
$\nu_{0}$, which we take as the K-band ($23~\rm{GHz})$.

The results for the synchrotron emission generally indicate
a steep spectral index consistent with a value in excess of
$\beta_s=3.0$. Curiously, the index is flatter for the $Kp0$ mask,
particularly when using the $g$-function.  The amplitudes, however,
fall with decreasing sky coverage -- this is generally true for all of
the emission components, suggesting that there remains within the
$Kp2$ to $Kp0$ transition region emission that differs in nature from
genuine high latitude emission.  It is also interesting to note that
the amplitude at 23 GHz implies a spectral index of $\sim$3 relative
to the 408 MHz emission, so that in general there are hints of
steepening at higher frequencies, as has been previously noted by the
\WMAP\ team.

\begin{figure}
\begin{centering}
\includegraphics[angle=90,width=1.05\textwidth]{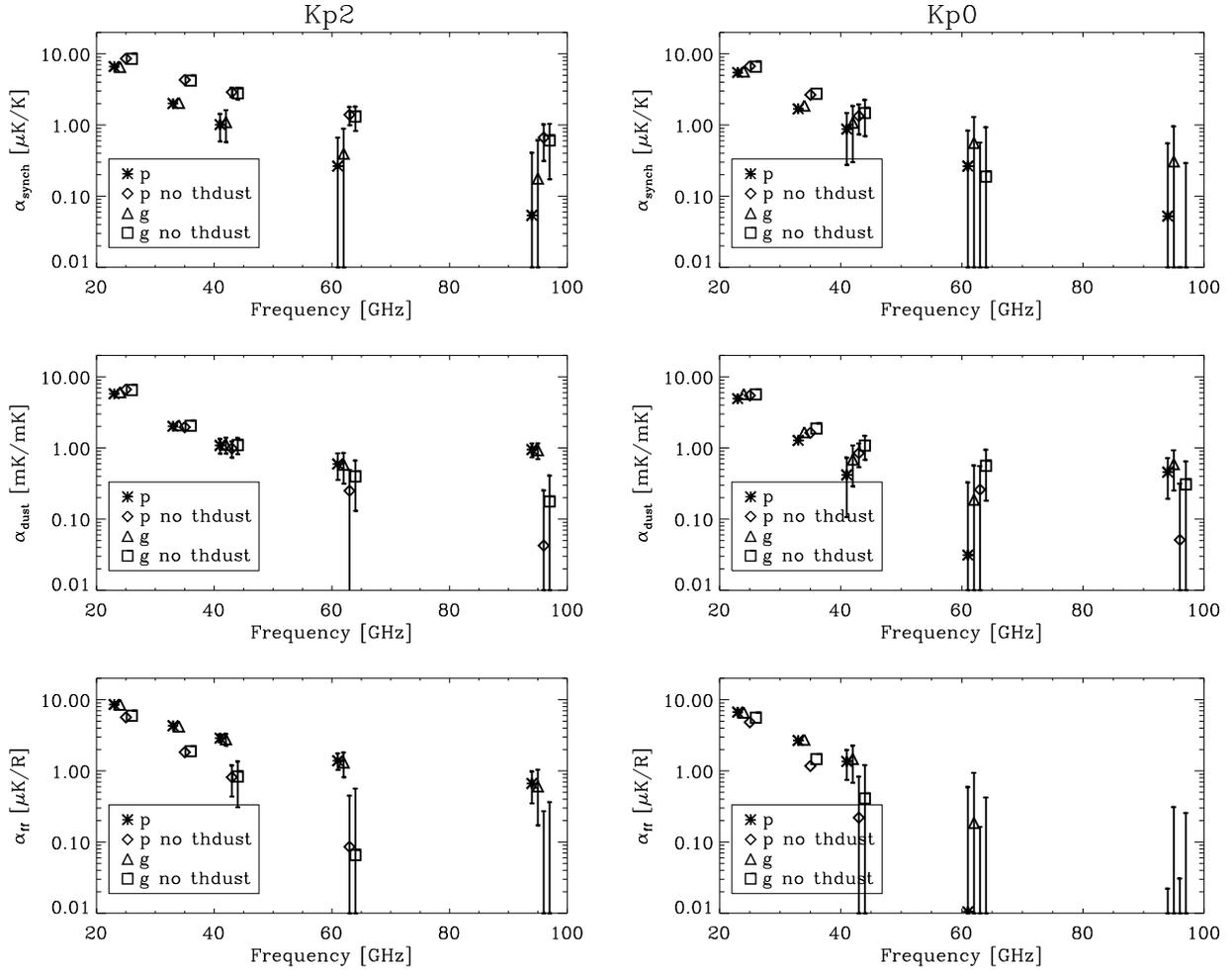}
\caption{\emph{Comparison between scaling factors obtained
from \WMAP\ data with and without the thermal dust emission, at the
resolution of $1^{\circ}$. At the lowest frequencies, the error
bars are smaller or comparable to the size of the plot symbols.
The only significant change is at W-band for the dust coefficients.
}}
\label{fig1}
\end{centering}
\end{figure}

Since the \WMAP\ frequency range does not allow a detailed study of the
spectral behaviour of the thermal dust component, we consider the 
spectral properties of the anomalous dust correlated emission.  Understanding the nature of
this component remains of great importance for the study of CMB
anisotropy and for motivating physical models of the emission.  We use
a single dust template to specify the morphology of both the thermal
dust emission and the anomalous dust correlated component. Thus, in
order to study the latter cleanly, we must apply a correction for the
thermal dust contribution.
Specifically, the scaling factors have been recomputed using {\fastica} after subtracting
the dust template from the data assuming that the FDS8 dust model (see
section \ref{dust}) provides an exact model of this emission
component.  Under this assumption a persistent correlation with the
dust template will be due to the anomalous component, spatially
correlated with the thermal dust, but with a different physical
nature. 
Figure~\ref{fig1} shows the
scaling factors for the dust emission without the thermal
component, compared with the values of the original analysis. In fact,
the values differ significantly for the dust coefficients only at $61$ and $94~\rm{GHz}$ as
might be expected since the thermal component falls off rapidly with
decreasing frequency.  Note also that whatever dust residuals remain
at the highest frequency are statistically insignificant, providing
strong support to the FDS8 model of thermal dust emission. 

\begin{table} 
\begin{center} 
\begin{tabular}{l | c | c| c | c| c | c }
\hline
\hline
&\multicolumn{2}{c}{\bf{Synchrotron}}
&\multicolumn{2}{c}{\bf{Anomalous Component}}&\multicolumn{2}{c}{\bf{Free-Free}}\\ 
\hline
&  $Kp2$&  $Kp0$ &  $Kp2$&  $Kp0$ &$Kp2$&  $Kp0$ \\
\hline 
&\multicolumn{6}{c}{{\fastica} -- function p}\\
\hline
$\beta$&$3.35\pm0.47$&$3.28\pm0.79$&$3.29\pm0.32$&$4.48\pm0.78$&$1.87\pm0.15$&$2.92\pm0.55$\\
$A_{norm}$&$6.64\pm0.43$&$5.46\pm0.62$&$5.68\pm0.26$&$4.85\pm0.33$&$8.56\pm0.38$&$6.79\pm0.63$ \\
\hline
&\multicolumn{6}{c}{{\fastica} -- function g}\\
\hline
$\beta$&$3.10\pm0.53$ &$2.82\pm0.80$&$3.35\pm0.35$&$4.14\pm0.75$&$1.92\pm0.22$&$2.72\pm0.65$\\
$A_{norm}$&$6.51\pm0.53$ &$5.62\pm0.80$&$5.98\pm0.29$&$5.63\pm0.41$&$8.51\pm0.52$&$6.68\pm0.81$\\
\hline 
\hline
\end{tabular}
\caption{\emph{Spectral index $\beta$ and normalisation
factor $A_{norm}$ obtained fitting values of the coupling coefficients for
synchrotron, the anomalous component of dust and free-free emission, 
with different masks. 
}}
\label{beta_1deg}
\end{center}
\end{table}

The values of the spectral index $\beta_a$ of the anomalous dust
component determined on the $Kp2$ sky coverage are 
somewhat steeper than
the value of $2.85$ obtained by \citet{davies_etal_2006}.
However, they have
also noted spectral indices as steep as 3.8 in several dust dominated
regions at mid- to high-latitude. 
As with the synchrotron component, the
amplitude of the emission drops for the $Kp0$ mask, and consequently
the spectral index steepens noticeably.  One might speculate that this is
related to the differing properties of the anomalous component closer
to the Galactic plane flattening the index of the $Kp2$ results. In
the context of the spinning dust models proposed by
\citet{draine_etal_1998a} to explain the anomalous emission, the
behaviour may reflect the properties of the spinning dust in different
phases of the interstellar medium, eg. a greater or lesser admixture
of spinning dust in the warm ionised medium (WIM) depending on latitude.

Finally, we consider the properties of the free-free emission as
represented by the coupling coefficients determined relative to the
H$\alpha$ template. Indeed, before determining the specific spectral
behaviour, it is worth noting that the derived coefficients are not in
accord with expectations for the H$\alpha$ to free-free conversion
factor for which a value of 8000\,$K$ is conventionally adopted for the
thermal electron temperature. In fact, the K- and Ka-band results for
$Kp2$ sky coverage are more consistent with temperatures in the range
5000-6000\,$K$. Curiously, however, the coefficients at higher
frequencies are increasingly consistent with a higher temperature
value. In terms of a power-law fit for the frequency spectrum, this
corresponds to a notably flatter spectral index than the typical value
of 2.14 for free-free emission, with the best-fit value for the $p$
function {\fastica} fits some 2$\sigma$ away from this canonical
amplitude.  However, the results when a $Kp0$ mask is applied are
significantly different, favouring an even steeper slope than expected
for free-free, and with normalisation amplitudes corresponding to an
electron temperature below 4000\,$K$.  This latter behaviour was also
observed by \citet{davies_etal_2006} for 5 regions specifically
selected to be dominated by free-free emission.

In order to assess the consistency of our results with the
theoretically motivated spectral index as compared to the
apparently anomalous best-fit models, we have 
fitted the values of the coupling coefficients
by an idealised model for the free-free emission,
$A_{norm}(\nu/\nu_{0})^{-2.14}$, and evaluated the goodness-of-fit
of the model.
The results are
consistent with the $\beta_{ff}=2.14$ model, even in the case of the $Kp0$ mask. 
However, in a set of 1000 simulations in which the 
simulated free-free emission followed the theoretically motivated
spectrum, no cases were found for which there were such large
changes in the fitted spectral slopes determined for $Kp2$ or $Kp0$
sky coverage as those observed with the data. This could reflect the ideal nature of the simulated
emission, but irrespective of this, although our results are statistically
consistent with the expected free-free scaling with frequency,
there are inconsistencies that are difficult to reconcile.

\begin{table} 
\begin{center} 
\begin{tabular}{l | c | c| c | c} 
\hline
\hline
&\multicolumn{4}{c}{\bf{Free-Free - Intensity}}\\
\hline
&\multicolumn{2}{c}{{\fastica} -- function p}&\multicolumn{2}{c}{{\fastica} -- function g}\\
\hline
&  $Kp2$  &  $Kp0$&  $Kp2$  &  $Kp0$\\
\hline
$\beta   $&$0.13\pm0.15$&$-0.92\pm0.55$&$0.08\pm0.22$&$-0.72\pm0.65$\\
$A_{norm}$&$0.14\pm0.01$&$ 0.11\pm0.01$&$0.14\pm0.01$&$ 0.11\pm0.01$\\
\hline
\hline
\end{tabular}
\caption{\emph{Spectral index $\beta$ and normalisation
factor $A_{norm}$ obtained fitting values of the intensity coupling coefficients for
free-free emission,  
with different masks. 
}}
\label{beta_ff_intensity}
\end{center}
\end{table}

\begin{figure}
\begin{centering}
\includegraphics[angle=90,width=0.65\textwidth]{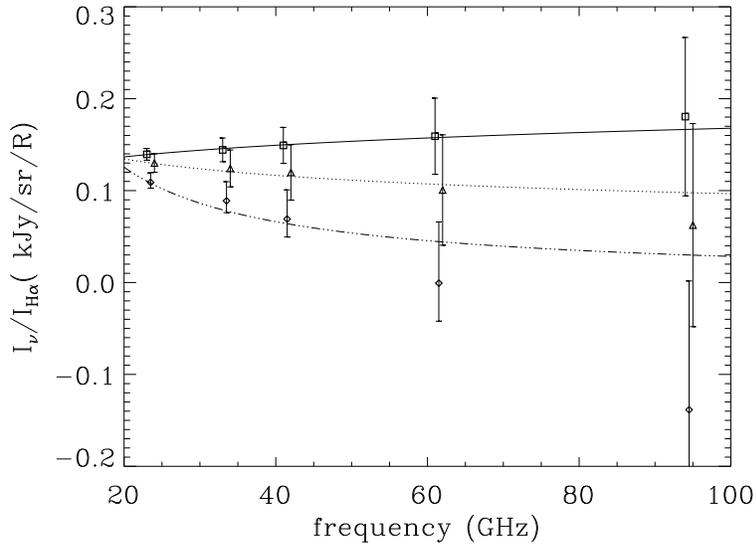}
\caption{\emph{The coupling coefficients in intensity units for 
free-free emission as traced by the H$\alpha$ template. Squares
represent the derived amplitudes for $Kp2$ sky coverage from the
{\fastica} $p$-function analysis, diamonds are for the corresponding
analysis on $Kp0$, whilst triangles show the results for a simple
$\chi^2$ analysis on $Kp2$. Best-fit curves are also shown.
The {\fastica} $Kp2$ results show an anomalous rising
spectrum for the free-free emissivity, whereas the $Kp0$ and 
$\chi^2$ $Kp2$ are steeper than expected, the former significantly so.
}}
\label{fig2}
\end{centering}
\end{figure}

Arguably then, the {\fastica} result for the $Kp2$ mask could be considered
an independent verification of the behaviour observed in
\citet{dobler_etal_2008a}. Motivated by their analysis, we augment our
assessment of the free-free spectral behaviour by fitting the
coefficients in intensity units. The results are presented in
Table~\ref{beta_ff_intensity} and Figure~\ref{fig2}.
\citet{dobler_etal_2008a} found that the H$\alpha$-correlated emission
showed a bump near to 50~GHz, although the exact behaviour was
sensitive to the type of CMB estimator that they pre-subtract from the
\WMAP\ data before fitting template amplitudes. In our analysis, no such
subtraction is necessary, since the coefficients are essentially
determined as part of the process of deriving the best CMB estimate
from the data. In fact, our $Kp2$ amplitudes show more extreme
behaviour at high frequencies, and indicate a systematically rising
spectrum, although the best-fit is also consistent within errors with
the slowly falling theoretically expected spectrum.  However, our
$Kp0$ results are significantly steeper than either the $Kp2$ or
expected slopes. Referring again to the results from the regional
analysis of \citet{davies_etal_2006}, there is a wide range of
spectral behaviour, with region 1 exhibiting a steeper rise in
frequency than seen here for $Kp2$, region 5 being somewhat consistent
with the observations here, but most regions falling off rapidly.
Interestingly, the simple $\chi^2$ analysis, for which a best-fit
spectral index of 0.2$\pm$0.3 is found, is very consistent with
theoretical expectations ($\sim$0.14).

The major difference between all of these results is how the method
treats the CMB contribution.  \citet{dobler_etal_2008a} subtract an
estimate of the CMB based on a variant of the \WMAP\ ILC method,
\citet{davies_etal_2006} explicitly include a CMB covariance term in
fitting the data directly with three templates, our $\chi^2$ analysis
simply ignores the CMB contribution in fitting the templates, whilst
the {\fastica} method is specifically attempting to construct the best
estimate of the CMB sky and computing template coefficients
accordingly. Clearly more work is needed to understand the relative
merits and problems in these approaches.

Finally, \citet{dobler_etal_2008b} interpret the results of both their
full-sky analysis and an assessment of the Gum nebula region in terms
of an enhanced emissivity due to a spinning dust contribution from the
WIM.  It is difficult to make a statement either
in support or against this conclusion based on our results, given the
large difference between the fits on $Kp2$ and $Kp0$, and the apparent
dependence on the CMB subtraction method. Perhaps it reflects the
different properties of the free-free emission close to the plane, or
the presence indeed of spinning dust in the WIM, or simply
cross-talk between different physical components that confuses
the spectral analysis. 
However, we remind the reader that our fits
for both $Kp2$ and $Kp0$ sky coverage are statistically
consistent with the expected scaling for free-free emission with
frequency, $\beta_{ff}=2.14$.

\subsection{Single Year Analysis}

\begin{figure}
\begin{centering}
\includegraphics[angle=90,width=1.05\textwidth]{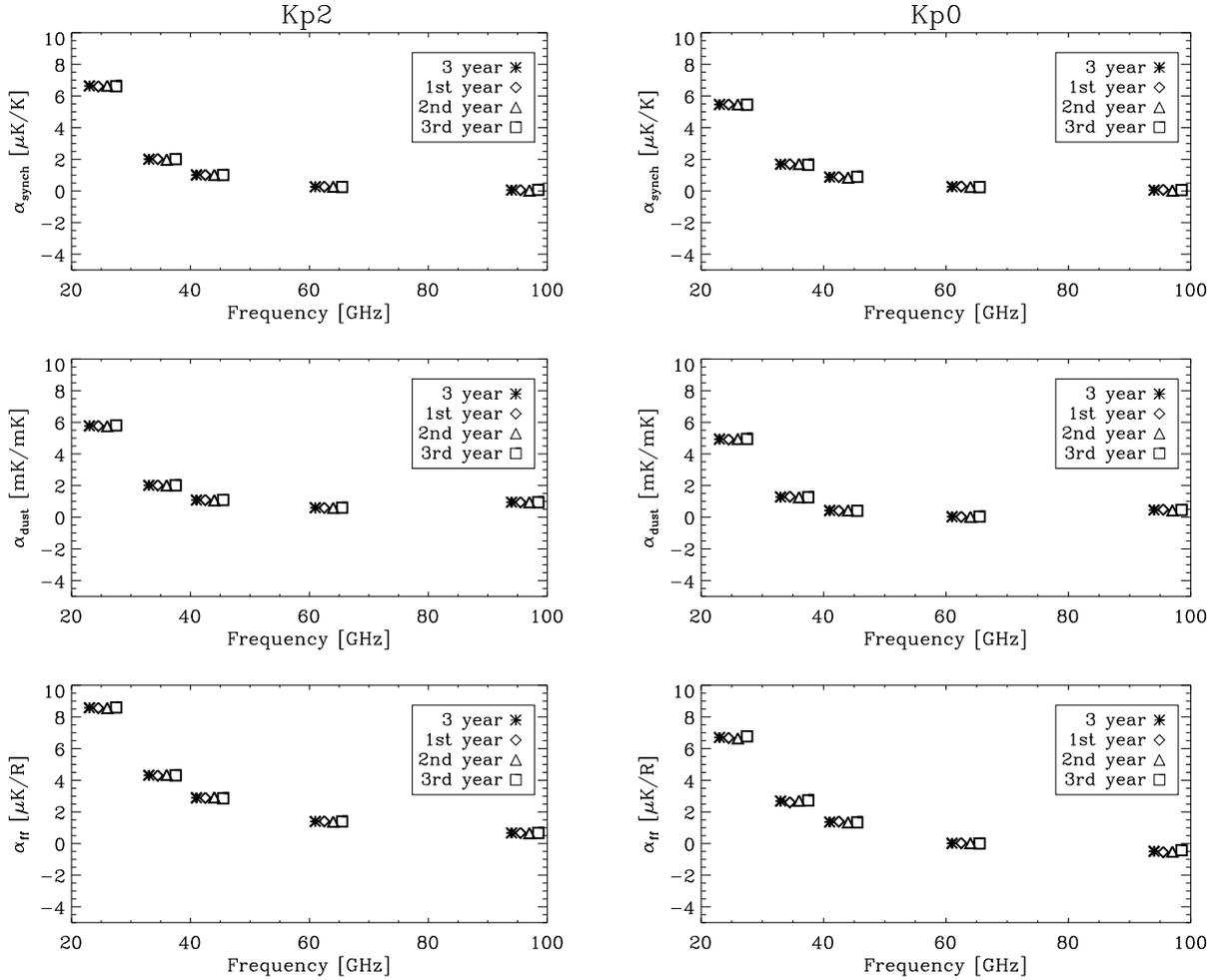}
\caption{\emph{Comparison between scaling factors determined
for the yearly \WMAP\ sky maps. The results for the 3-year \WMAP\ data
are also shown for reference. All results are determined using the
{\fastica} p-function.
}}
\label{fig3}
\end{centering}
\end{figure}

In order to test the stability of the coupling coefficients between
the foreground templates and the \WMAP\ data to potential systematics, 
eg. annual instrumental or calibration variations, 
we have performed the same {\fastica} analysis for the yearly sky maps.
These results are plotted in 
Figure~\ref{fig3}
together with the 3-year data coefficients. 
We find no evidence of significant annual variations in the coefficients.

\section{Analysis with the K-Ka map as synchrotron template.}
\label{K-Ka_analysis}

One of the limitations of the previous foreground analysis is that the
Haslam map used as template for the synchrotron emission is unlikely
to be completely representative of the foreground morphology at the
\WMAP\ frequencies due to spectral index variations across the sky.
Therefore, we repeated the analysis using the K-Ka map as an internal
template for the synchrotron emission. We then computed the
scaling factors for the Q, V and W channels.  The results are
shown in Table~\ref{tab_corr_and_error_bars_k-ka_1deg}.  As before,
the uncertainties of the coupling coefficients have been computed
from 1000 simulations. Once again, the $p$-function seem to be
statistically more robust than the corresponding $g$-function ones.

Table~\ref{tab_corr_and_error_bars_k-ka_1deg} also contains results
returned by the $\chi^2$ minimisation method, both our own and those
from \WMAP\ \citep{hinshaw_etal_2007}. For our analysis, we did not
impose any constraints on the spectral behaviour of the various
physical components. In contrast, \citet{hinshaw_etal_2007} explicitly
account for the free-free signal present in the K-Ka template, and
impose constraints on the free-free ($\beta_{ff}=2.14$) and thermal
dust ($\beta_{d}=2$) spectral indices.  The coefficients derived for
the K-Ka and dust templates are in excellent agreement for all three
methods. The free-free values, however, are notably different in all
three cases. The difference to the \WMAP\ results could, at least in
part, be due to their application of a large correction for dust
absorption, although comparing coefficients at all three frequencies
the explanation is clearly more complex.  A more useful observation in
regard to the free-free coefficients is that they are universally
lower than in the corresponding Haslam case for the Q-band but almost
identical for W-band. This directly reflects the fact that the K-Ka
template contains not only synchrotron but also free-free emission.
The latter also indicates that the {\fastica} results continue to prefer a
flatter slope for the free-free emission than expected from theory,
and as derived by the $\chi^2$ analysis.

The most interesting results are related to the amplitudes of the
coupling coefficients for the dust. In particular, note the low
value of the dust contamination in the Q-band in contrast with
previous findings for the Haslam case.  This could be interpreted as
indicating the absence of an anomalous dust correlated component, in
apparent contradiction to previous results. However, it is also
consistent with a picture in which the K-Ka template contains
contributions not only from the synchrotron and free-free emission,
but also the anomalous component, which the Haslam map certainly does
not. At W-band and for the $Kp2$ sky coverage, the K-Ka contribution is
essentially zero, so that there is no effective contribution of this
embedded anomalous component, and the dust coefficients should be
consistent with the thermal dust emission. Indeed, there is again
strong support for the FDS model of the thermal dust emission,
although for the $Kp0$ cut the amplitude drops as before.

In order to understand whether there is a self-consistent picture with
the Haslam results, we can perform some simple numerical
comparisons. If we assume that the \WMAP\ data contains foregrounds
comprised of synchrotron, free-free and anomalous dust contributions
as described in Table~\ref{table_1deg_coeffs_haslam}, then we can
infer the extent of these components present in the K-Ka template, and
then predict the amplitudes expected for the free-free and dust
contributions in Table~\ref{tab_corr_and_error_bars_k-ka_1deg}
allowing for that fraction contained in the K-Ka map. The consistency
is excellent for both $p$ and $g$ function results on the $Kp2$ sky
cut, and certainly good for the $Kp0$ coverage.  We have also made a
more detailed study using simulations, with the same results. Of
course, this does not imply that the K-Ka template is not a
significant improvement on the Haslam map, but that the gross features
of the two foreground models are in agreement. The quality of the two
synchrotron templates will be considered further in the following sections.

\begin{table}
\begin{center} 
\begin{tabular}{l | cccccc } 
\hline
\hline
& \multicolumn{2}{c}{\bf{Synchrotron}} &  \multicolumn{2}{c}{\bf{Dust}} &  \multicolumn{2}{c}{\bf{Free-free}}  \\ 
\hline
&$Kp2$&$Kp0$& $Kp2$&$Kp0$&$Kp2$&$Kp0$\\
\hline 
&\multicolumn{6}{c}{\bf{{\fastica} - function p }}\\
\hline
$Q$   & $0.24\pm0.04 $&  $0.21\pm0.07$  & $0.17\pm0.32$& $-0.13\pm0.41$ & $1.92\pm0.42$& $0.48\pm0.70$ \\
$V$   & $0.05\pm0.04 $&  $0.03\pm0.06$  & $0.50\pm0.29 $& $0.40\pm0.30$& $1.19\pm0.40$& $-0.16\pm0.66$\\
$W$   & $-0.01\pm0.03$&  $-0.04\pm0.05$ & $1.00\pm0.26$& $0.77\pm0.34$  & $0.73\pm0.35$& $-0.41\pm0.58$ \\
\hline 
&\multicolumn{6}{c}{\bf{{\fastica} - function g }}\\
\hline 
$Q$    & $0.24\pm0.09$& $0.16\pm0.15$   & $0.16\pm0.52$& $0.03\pm0.80$  & $1.76\pm0.66$& $0.84\pm0.98$ \\
$V$    & $0.06\pm0.08$& $-0.004\pm0.14$ & $0.23\pm0.76$& $0.39\pm0.50$ & $1.05\pm0.63$ & $0.16\pm0.93$\\
$W$    & $0.01\pm0.072$& $-0.05\pm0.12$& $0.94\pm0.44$& $0.80\pm0.66$ & $0.56\pm0.54$& $-0.20\pm0.81$\\
\hline
\hline
&\multicolumn{6}{c}{\bf{$\chi^2$ analysis (no constraints)}} \\
\hline
$Q $&$0.21\pm0.09$&&$0.42\pm0.50 $&&$1.44\pm0.77$&\\
$V $&$0.03\pm0.08$&&$0.57\pm0.46 $&&$0.75\pm0.70$&\\
$W $&$-0.01\pm0.06$&&$1.10\pm0.36 $&&$0.29\pm0.54$&\\
\hline
 &\multicolumn{6}{c}{\bf{Hinshaw et al.}} \\ 
\hline 
$Q$  &$0.23$&&$0.19$&&$0.99$&\\
$V$  &$0.05$&&$0.41$&&$0.63$&\\
$W$  &$0.00$&&$0.98$&&$0.32$&\\
\hline
\hline
\end{tabular}
\caption{\emph{Values of coupling coefficients in antenna
temperature units between the 3-year \WMAP\ Q,V, and W-band data and 
three foreground emission templates 
at $1^{\circ}$ resolution. The K-Ka map is
adopted as the synchrotron template. 
The {\fastica} analysis
is performed using the two non-linear functions $p$ and $g$ 
and for the $Kp2$ and $Kp0$ masks.
The corresponding $Kp2$ results for a simple $\chi^2$ analysis
are provided for comparison. 
In addition, we provide the band-averaged values 
from the \citet{hinshaw_etal_2007} fits
performed with constraints imposed on the 
free-free and thermal dust spectral indices.
Note that the \WMAP\ team 
used an H$\alpha$ template corrected for dust absorption 
assuming $f_d=0.5$ for their analysis.
The units are mK/mK for synchrotron, mK/mK
for dust and $\mu$K/R for free-free emission respectively. }}
\label{tab_corr_and_error_bars_k-ka_1deg}
\end{center}
\end{table}

\subsection{Spectral properties of foreground emissions.}
\label{k_ka_spectra}

A study of the spectral index of the returned K-Ka coefficients would
be unphysical, or at least difficult to interpret, since the template
is a mix of synchrotron, free-free and anomalous dust emission.
Moreover, the interpretation of the spectral behaviour of the returned
dust and free-free coefficients is similarly problematic, requiring a
detailed understanding of the fraction of the signal associated with
the K-Ka template. However, it is worth reiterating, as shown above,
that the template fits are in excellent agreement with those in which
the Haslam map is used as the synchrotron template, and therefore the
spectral behaviour of the components can be interpreted in the same
manner.

\section{EVALUATION OF THE Q-, V- AND W-BAND SKY MAPS AFTER FOREGROUND
CLEANING}

One of the main aims of the foreground analysis performed thus far
with {\fastica} is to promote the possibility of cleaning the \WMAP\
data sufficiently to allow its use for cosmological purposes.
Therefore, adopting the coupling coefficients derived for various
foreground templates, we have studied \WMAP\ sky maps cleaned of the
Galactic emission. 

\begin{figure}
\begin{centering}
\includegraphics[angle=90, width=1.\textwidth]{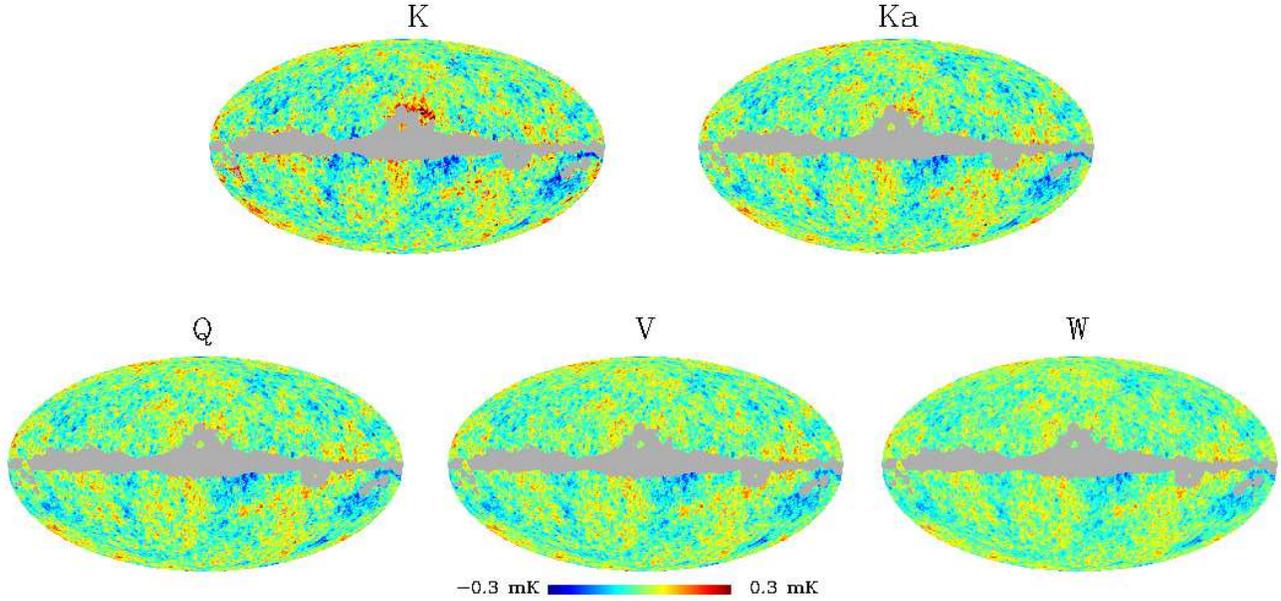}
\caption{\emph{WMAP data for K-, Ka-, Q-, V- and W-bands cleaned by
subtracting foreground templates scaled by the coupling
coefficients determined by {\fastica} with the $p$-function for the
$Kp2$ sky coverage. 
Here, the synchrotron template is the Haslam map. 
The maps are shown in a conventional mollweide
projection in a Galactic frame-of-reference , with the north pole
at the top of the image and the Galactic Center in the middle 
with longitude increasing to the left.
The regions in grey
correspond to the bright Galactic emission and point sources excised
from the analysis by the $Kp2$ mask.
}}
\label{fig4}
\end{centering}
\end{figure}

\begin{figure}
\begin{centering}
\includegraphics[angle=90, width=1.\textwidth]{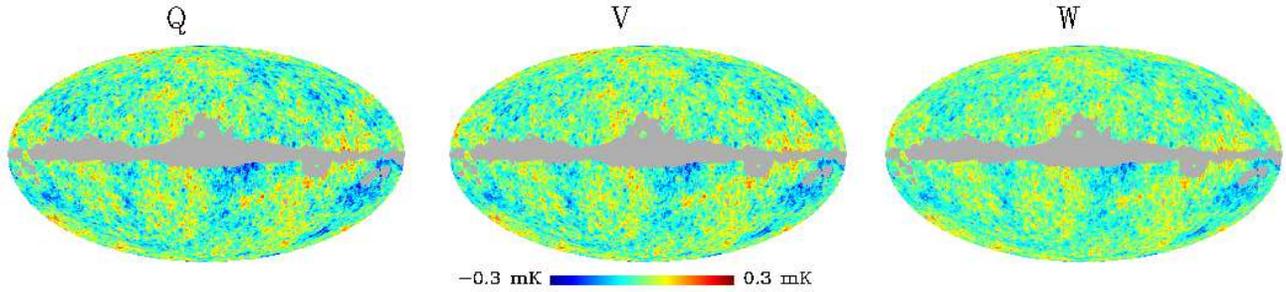}
\caption{\emph{WMAP data for Q-, V- and W-bands cleaned by
subtracting foreground templates scaled by the coupling
coefficients determined by {\fastica} with the $p$-function.
Here, the synchrotron template is the K-Ka map. 
}}
\label{fig5}
\end{centering}
\end{figure}

Figure~\ref{fig4} shows the residual sky
signal of the five \WMAP\ channels that have been cleaned using the
Haslam data as the synchrotron template.  We show only the results
based on the $p$-function analysis, since there is little visual
difference relative to the $g$-function.  There are clearly residuals
close to the Galactic plane for the K and Ka sky maps, particularly
near the edge of the $Kp2$ mask in the vicinity of the Galactic
Centre. This excess foreground emission falls off rapidly with
frequency, and only a hint can be seen in the Q-band
data. Nevertheless, the dominant frequency-independent CMB structure
is clear at high latitudes.  Figure~\ref{fig5}
shows the equivalent results for the Q-, V- and W-bands when the K-Ka
synchrotron template is adopted. There is little visual evidence of
residual foregrounds even close to the masked regions.

\begin{figure}
\begin{centering}
\includegraphics[angle=90,width=1.\textwidth]{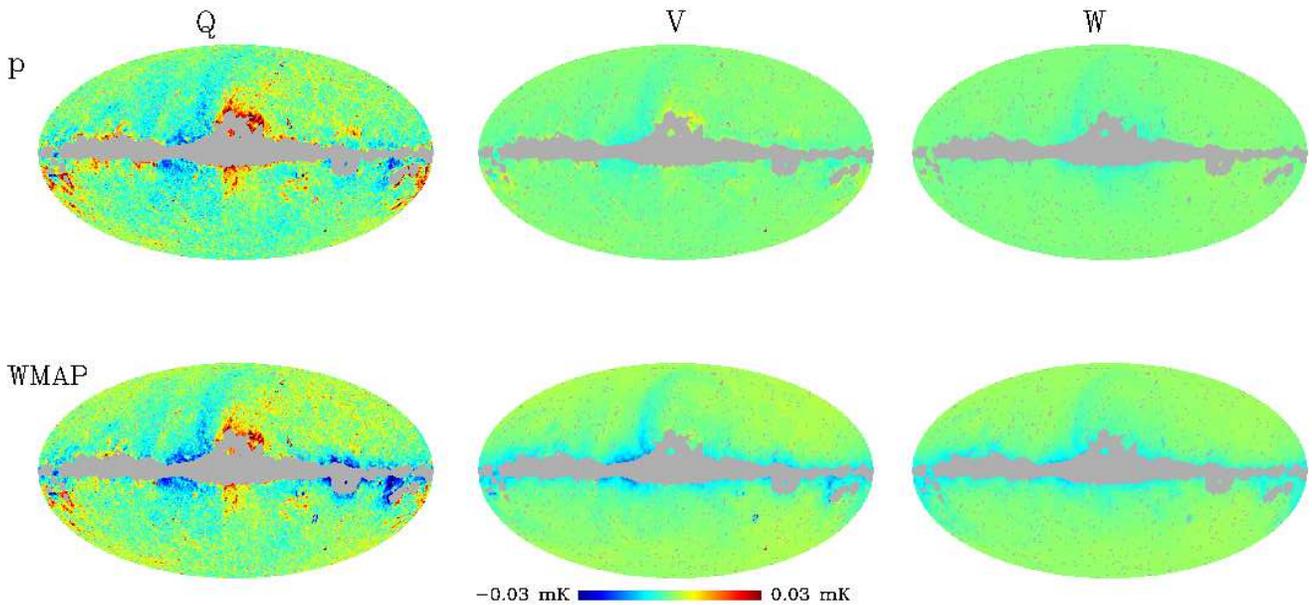}
\caption{\emph{Difference between the cleaned maps using the Haslam
map and the K-Ka map at $41$, $61$ and $94~\rm{GHz}$.
The top row shows the difference when the {\fastica} $p$-function
coefficients are used, the lower line shows the equivalent difference
maps from when applying the \WMAP\ first-year and three-year foregrounds
corrections.}}
\label{fig6}
\end{centering}
\end{figure}

As an interesting comparison, we have investigated the difference
between the cleaned maps obtained using either the Haslam or the K-Ka
synchrotron templates, scaled by the coupling coefficients determined
either by our {\fastica} analysis or by
the \WMAP\ team.  Figure~\ref{fig6} shows the
results for the Q-, V- and W-bands (in neither the first nor second
data release did \WMAP\ provide cleaned K- or Ka-data, presumably
because they do not utilise them for cosmological analysis).
The main difference apparent at Q-band is the presence of a bright
structure around the mask concentrated near the Galactic Centre both
above and below the plane. This is a clear delineation of the K-band
residual noted above when using the Haslam template, and is most
easily interpreted as Galactic emission that is not traced by the
Haslam synchrotron template.  It therefore supports the \WMAP\ assertion
that adopting the K-Ka map as a template provides a more complete
model of the foreground emission at low frequency where other
contributions beyond the synchrotron and free-free emission may exist.
This structure is seen clearly in both the {\fastica} and \WMAP\
residual plots. Also seen is evidence of a negative residual
associated with a well-known feature in the low frequency template -- the North Polar
Spur -- plus other residuals around the edge of the Galactic plane
part of the $Kp2$ mask, although whether these are more associated
with the Haslam or K-Ka analysis is unclear.

The residuals at V- and W-bands show much fainter features, and
particularly for the {\fastica} analysis. Indeed, in this case the
plot indicates that the total foreground level determined using either
the Haslam or K-Ka maps, plus the usual tracers of free-free and dust
emission, are in very good agreement. This is consistent with the
discussion in the previous section. For the data corrected using the
\WMAP\ template coefficients, the residuals seem to be larger in all
cases. 

The comparison here cannot easily be used to infer whether the
{\fastica} template fits are to be preferred to the \WMAP\ ones, but we
will attempt to address this issue in the next section.  What should
be clear is that it is always good practise for a cosmological
analysis to study the cleaned maps at different frequencies and test
for evidence of a frequency dependence that is the signature of
foreground residuals.
 
\begin{figure*}
\begin{centering}
\includegraphics[width=0.45\textwidth]{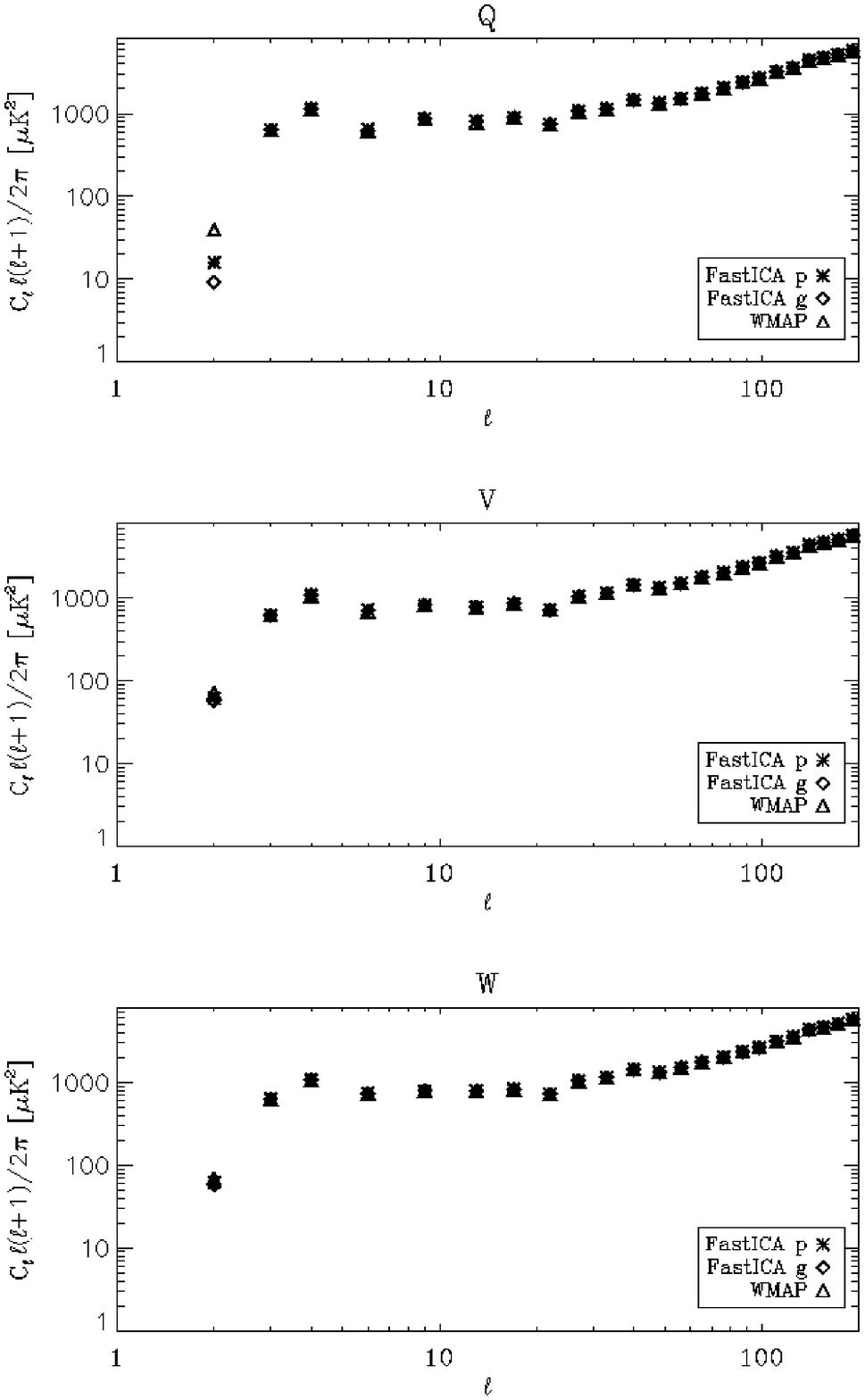}\qquad\qquad
\includegraphics[width=0.45\textwidth]{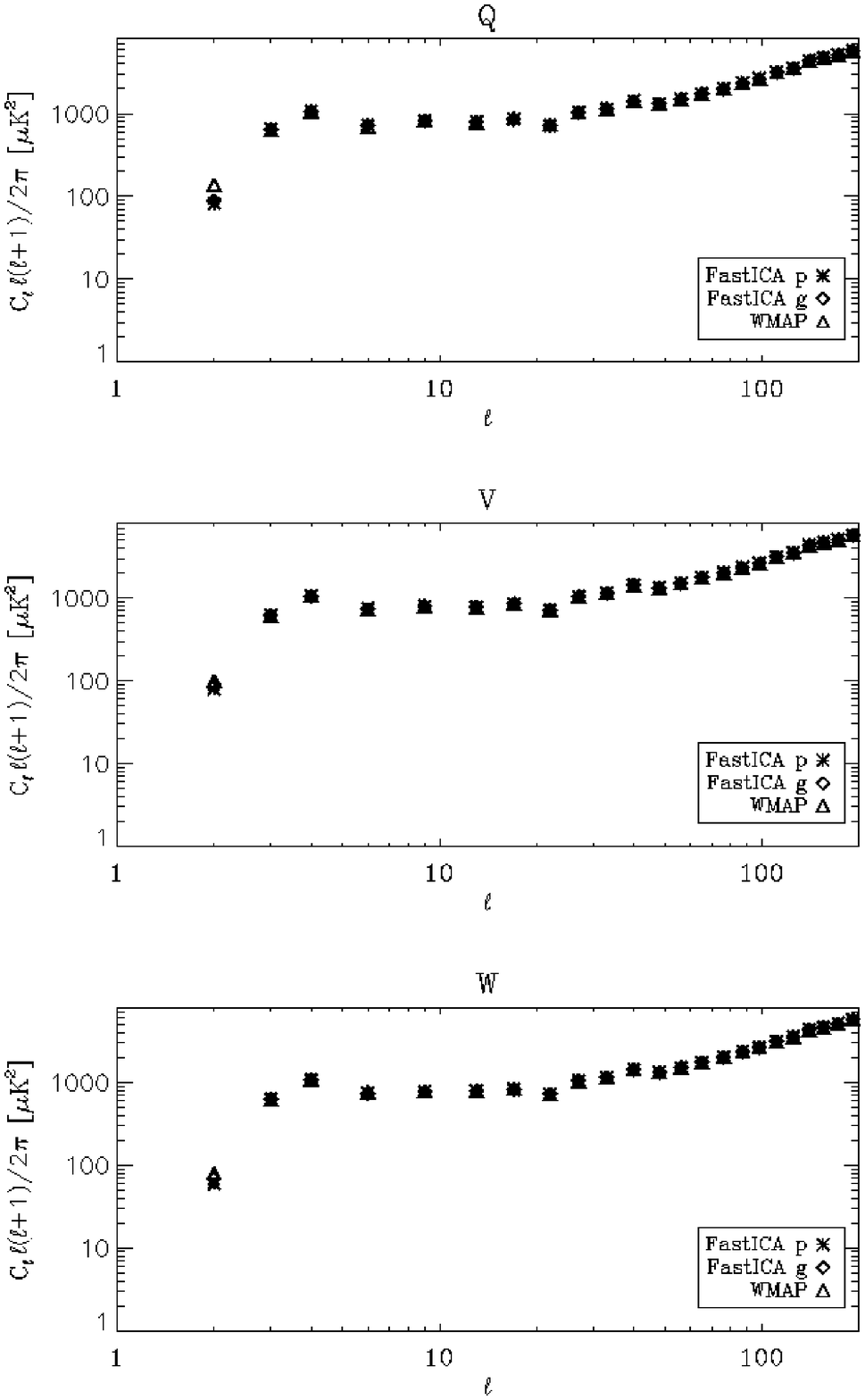}
\caption{\emph{Left: Binned power spectra of the Q-, V- and W-maps cleaned from the foreground emission.
Their contamination is described by the coupling
coefficients determined by {\fastica} with both
the $p$- and $g$-functions (asterisk and diamond respectively) 
and by the \WMAP\ analysis (triangle). The left figure corresponds to
the case when the Haslam map is used as the synchrotron template, the
right figure to the corresponding analysis using K-Ka for the
low-frequency emission.}}
\label{fig7}
\end{centering}
\end{figure*}

In fact, we do exactly that by examining the power spectra of the
cleaned maps determined using the MASTER algorithm \citep{hivon_etal_2002}.  
Figure \ref{fig7}
indicates that the maps are essentially insensitive to the residual
foreground features that remain in the data due to the different
fitting methods and templates, although the quadrupole amplitude
is generally suppressed, and
notably so when using the Haslam template to trace synchrotron
emission. Thus the cleaned maps, particularly at
V- and W-bands, can be considered adequate for cosmological analyses,
at least when based on the angular power spectrum.

\section{AN \lq ITERATIVE' BLIND COMPONENT SEPARATION STUDY USING \WMAP\
DATA PRE-CLEANED USING TEMPLATES}
\label{cleaned_data_analysis}

As demonstrated in the previous section, the \WMAP\ data cleaned by
foreground templates still show evidence of contamination due to
either the inability of the templates to describe real spectral
variations in the foreground emission on the sky, or other emission
mechanisms not identified with the standard Galactic
components/templates.  Inspired by the analysis of
\citet{patanchon_etal_2005}, we have performed an \lq iterative' blind
analysis on the \WMAP\ data, \emph{after} pre-cleaning with the
foreground templates, in order to study the residuals in more detail.

\begin{figure}
\begin{centering}
\includegraphics[angle=90,width=1.\textwidth]{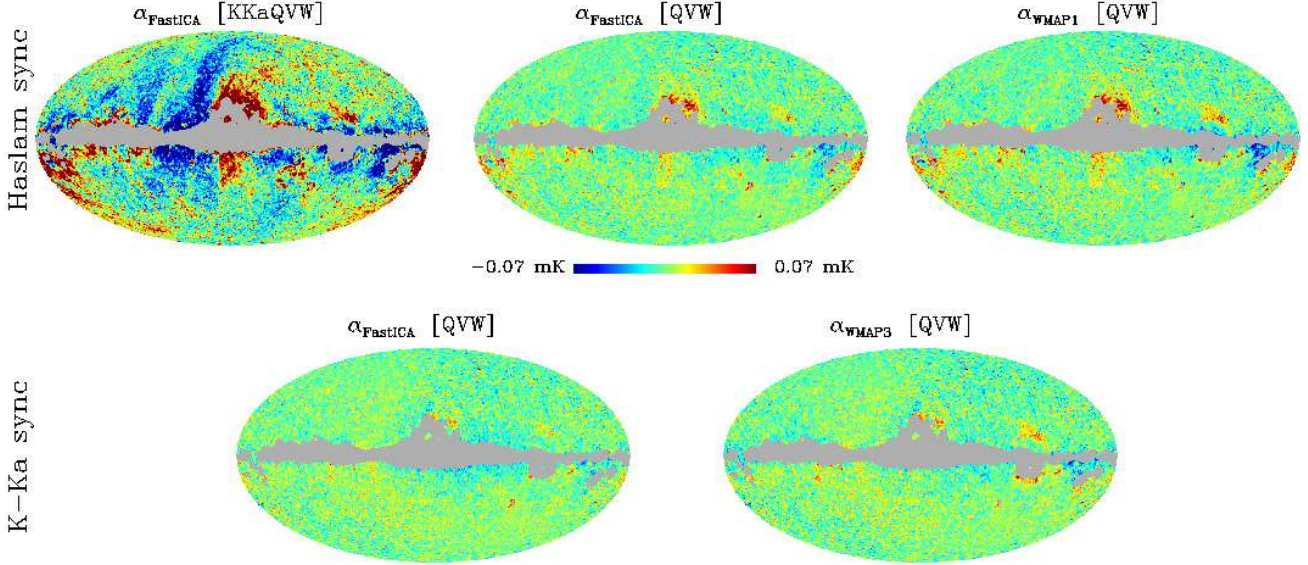}
\caption{\emph{Maps of the residual components obtained by {\fastica}
when applied to the \WMAP\ data after pre-cleaning using template fits
with coupling coefficients $\alpha$ derived either using a {\fastica} analysis 
with the $p$-function and the $Kp2$ mask or by the \WMAP\ team in
\citet{bennett_etal_2003} (WMAP1) and \citet{hinshaw_etal_2007} (WMAP3).
Several cases have been considered:
i) input maps K-, Ka-, Q-, V- and W-bands plus Haslam synchrotron template,
{\fastica} coefficients;
ii) input maps Q-, V- and W-bands plus Haslam synchrotron template,
{\fastica} coefficients;
iii) input maps Q-, V- and W-bands plus Haslam synchrotron template,
\WMAP\ constrained fit coefficients;
iv) input maps Q-, V- and W-bands plus K-Ka synchrotron template,
{\fastica} coefficients;
v) input maps Q-, V- and W-bands plus K-Ka synchrotron template,
\WMAP\ constrained fit coefficients.
}}
\label{fig8}
\end{centering}
\end{figure}

Various combinations of the cleaned \WMAP\ channels are used as input to
the {\fastica} algorithm. This then returns a number of components
equal to the number of the input sky maps.  One of these is identified
as a CMB map, cleaned as far as possible from any \emph{residual}
emission (see section~\ref{subsec:cmbresids}). Without exception, at
most one of the other returned components
can be clearly indentified with Galactic residuals, the remainder
generally being some combination of anisotropic noise (reflecting the
underlying observation pattern of the \WMAP\ satellite) and weak
residual dipole emission.
Figure~\ref{fig8} shows maps of the candidate foreground
residuals for a number of \WMAP\ and template input permutations,
as described in the figure caption and discussed in section~\ref{subsec:fgresids}

\subsection{Foreground residuals}
\label{subsec:fgresids}

The upper row of Figure~\ref{fig8} shows the output maps
from {\fastica} that we identify with foreground residuals when using
the Haslam template for synchrotron emission. 

The leftmost figure presents results from analysing the five \WMAP\ frequency bands
after cleaning using the coefficients previously derived in this paper
(see Table~\ref{table_1deg_coeffs_haslam}).  There are some bright
features along the Galactic plane near to the Galactic Centre and the
presence of a structure related to the North Polar Spur. These are a
clear indication of the deficiencies of the Haslam map as a
synchrotron template, particularly for cleaning the K- and Ka-band
data, with which the residuals are predominantly associated. Some of
the positive and negative features presumably reflect either under-
or over-subtration of the true foreground signal by the templates
as a consequence of spectral variations.

The middle figure shows the equivalent analysis using just the
Q-, V- and W-band pre-cleaned sky maps. The residuals are 
significantly reduced, and no evidence of the North Polar Spur
is now observed. The dominant feature, as before, is the structure 
close to the Galactic centre.
This was originally identified and referred to as the \lq free-free
haze' by \citet{finkbeiner_2004} (although its origin as free-free
emission is unlikely and it is now referred to as the \lq \WMAP\ haze'), 
and was already visible
in the foreground residuals plot of \citet{bennett_etal_2003}.
Moreover, the map is strikingly (though perhaps unsurprisingly) 
similar to the map obtained by \citet{patanchon_etal_2005}, using Spectral
Matching Independent Component Analysis (SMICA). They have associated these
residual structures with the Ophiuchus complex, 
the Gum nebula, the Orion-Eridanus bubble and the Taurus region.
The right-hand plot is the same analysis when the three input maps
are cleaned using the \WMAP\ foreground model of
\citet{bennett_etal_2003}. The residual features are in excellent
agreement with those from the middle plot, although the amplitudes
are slightly larger, suggesting that the {\fastica} template coefficients
provide a modestly better cleaning of the \WMAP\ frequency data.

The residual maps from input data cleaned with the K-Ka synchrotron model
are shown in the second row of Figure~\ref{fig8}.
The maps can be compared directly to the equivalent Haslam cases 
in the row above. It is particularly striking that the
residuals are substantially suppressed when using the K-Ka template,
especially close to the Galactic Centre. This reinforces the latter's
superiority for foreground separation purposes, although the fact that
it is a mixture of physical emission components makes the
interpretation of the results more difficult. As before, the
residuals are larger when using the \WMAP\ template coefficients,
notably so near the Gum nebula -- a source of strong free-free
emission.
 
\subsection{CMB component}
\label{subsec:cmbresids}

\begin{figure}
\begin{centering}
\includegraphics[angle=90.,width=1.\textwidth]{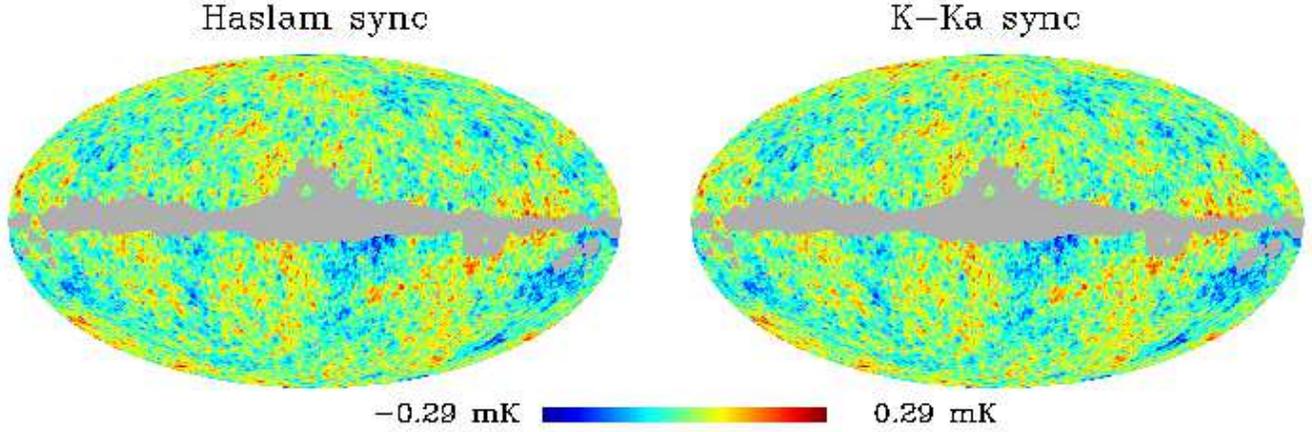}
\caption{\emph{Map of the CMB component obtained by {\fastica} with
the $p$-function using \WMAP\ Q-, V- and W-band data pre-cleaned using
coupling coefficients estimated
with the $Kp2$ mask, and either the Haslam map as the synchrotron
template (left) or the K-Ka map (right). The temperature is measured in mK.
}}
\label{fig9}
\end{centering}
\end{figure}

\begin{figure*}
\begin{centering}
\includegraphics[width=0.7\textwidth]{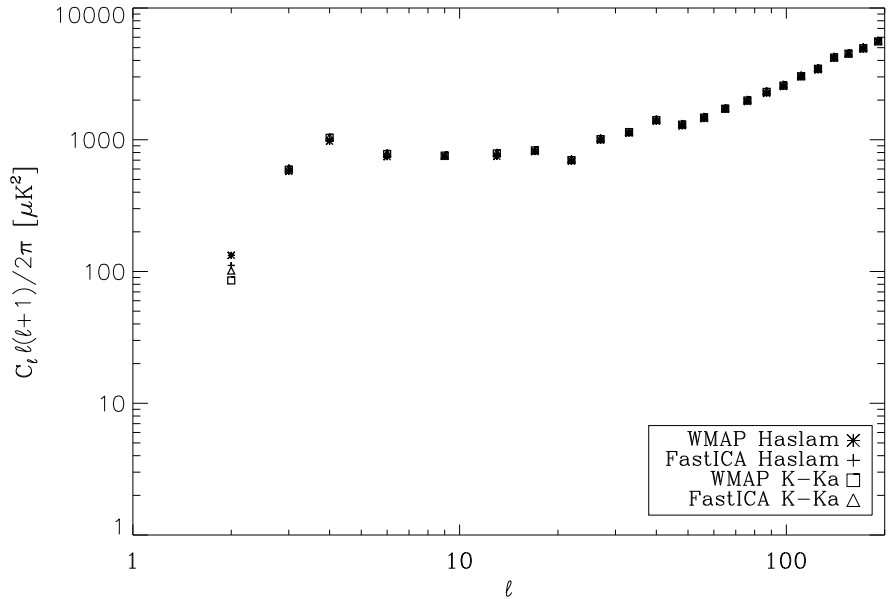}
\caption{\emph{Power spectra of the returned CMB components applying
{\fastica} to the set of Q-, V- and W-band data from the 
foreground emissions, using various combinations of foreground
templates and fit coefficients. }}
\label{fig10}
\end{centering}
\end{figure*}

In order to quantify
the improvement connected to the 
iterative processing step, 
we have studied
the CMB components returned by {\fastica} when working
on the data pre-cleaned with a template analysis stage. They are
very similar, as shown in Figure~\ref{fig9}.
This is also confirmed by the plot of the power spectrum for 
the CMB map derived from the different cases of the
foreground analysis. 
Figure~\ref{fig10} compares the binned power spectrum 
of the CMB maps returned from the Q-, V- and W-band data cleaned
using the coupling coefficients of the {\fastica} analysis or
the \WMAP\ values, with either the Haslam or K-Ka maps as synchrotron
template. The discrepancy is less than $150~\mu K^2$.

Given this, one might argue that the iterative step is unnecessary for
\WMAP\ analysis, but we fully expect that it will be essential for
studying the \emph{Planck} data in detail. Moreover, in this era of
precision cosmology, it is the deviations from the theoretically
motivated model spectra which are of interest and the search for
evidence of primordial non-Gaussianity requires that false detections
due to foreground residuals be eliminated. {\fastica} can play a role
in the initial cleaning of the data, and then in defining regions on
the sky that merit special treatment, be it simply masking them from
future cosmological analysis due to the apparenty complex nature of
the foregrounds there, or the re-application of the component
separation techniques on the regions to improve the final CMB sky
estimate.

\section{DISCUSSION}
\label{discussion}

In this paper, we have utilised a blind component separation algorithm
-- {\fastica} -- to address the issue of the Galactic foreground
emission in the \WMAP\ 3-year data. In particular, we have adapted the
technique to allow three Galactic foreground templates (that trace
synchrotron, free-free and thermal dust emission) to be fitted to the
data on a frequency-by-frequency basis.  Specifically, we use the
\citet{finkbeiner_2003} H$\alpha$-map as a template for the free-free
emission, the \citet{finkbeiner_etal_1999} FDS8 model for thermal
dust emission, and for synchrotron emission two alternatives -- the
408~MHz radio continuum all-sky map of \citet{haslam_etal_1982} as
utilised in the first year \WMAP\ analysis, or the difference of the
\WMAP\ K- and Ka-band data preferred in their three year analysis.

Detailed simulations indicate that the algorithm behaves in an
unbiassed way only for two ($p$, $g$) of the three non-linear
functions typically applied by {\fastica}. In these cases, the
$p$-function also appears to be marginally preferred as a consequence
of its smaller error bars, both relative to the $g$-function analysis
and the simple $\chi^2$ method that is usually adopted for template
fitting.

When applied to the \WMAP\ data, we have quantified the foreground
contamination in term of coupling coefficients between the data and
the foreground templates. The derived values are sensitive to both the
extension of the mask applied to exclude the Galactic plane and on the
non-linear function used by the algorithm.  The mask dependence is
likely to reflect genuine variations in the spectral behaviour of the
foregrounds with sky location, as seen in previous local studies of
template fits by \citet{davies_etal_2006}. The non-linear function
dependence is plausibly linked to sensitivity to the actual
statistical nature of the foreground components (as represented by the
templates), and suggests that it may be useful to study more
appropriate approximations to the neg-entropy.

We have considered the spectral behaviour of the derived scaling
factors when the \citet{haslam_etal_1982} data is used as the
synchrotron template.  We evaluated the spectral index for the
synchrotron emission, the anomalous dust-correlated component, and the
free-free emission.  In the first two cases, we found steeper, though
statistically consistent, spectral behaviour as compared to previous
analysis, eg. \citet{davies_etal_2006}.  For the free-free emission,
we confirmed previous observations about inconsistencies in the fitted
scaling to convert the H$\alpha$ emissivity to a free-free
contribution at K- and Ka-bands.  Specifically, the derived factors
correspond to thermal electron temperatures of $\sim 4000\, K$
compared to the expected value of $8000\, K$.

In addition, we find independent verification of a flat spectral
dependence in the free-free emissivity as reported by
\citet{dobler_etal_2008a}. However, this behaviour is seen only for
the $Kp2$ sky coverage. With the $Kp0$ mask, the behaviour is even
steeper than the expected $\nu^{-2.14}$ dependence, suggesting that
the anomalous free-free behaviour is associated with bright structures
close the Galactic plane.

We have also proposed that the {\fastica} algorithm can be applied \lq
iteratively', that is, used to analyse a set of multi-frequency maps
pre-cleaned using templates for which the 
coefficients have themselves been derived by a {\fastica} analysis. This is
particularly useful in order to determine the presence of residual
foregrounds that arise either due to a mismatch between their spectral
dependence and the average high-latitude value determined by the template fits, or
because they are new components that are not traced by the adopted
templates. In this manner, we have confirmed the existence of a
component spatially distributed along the Galactic plane, with pronounced
emission near the Galactic center. 
This is the emission previously noted in the SMICA analysis
of \citet{patanchon_etal_2005}, and is the \lq \WMAP\
haze' of \citet{finkbeiner_2004}.
However, the amplitude and extent of the emission is less when
the K-Ka map is adopted as the synchrotron template. This is not
surprising if the foreground component is a genuinely new 
contribution to the sky emission that is untraced by the Haslam
data. In this context, the K-Ka map must be considered the better
template for cleaning the \WMAP\ data from foregrounds. However,
the interpretation of the results in terms of Galactic emission
components is complicated by the fact that 
the K-Ka template contains a mixture of several, physically distinct,
emission mechanisms.

Finally, the analysis that we have performed with the {\fastica}
algorithm is based on the
unrealistic hypothesis that the spectral behaviour of the various foreground
emission components is unchanging over the sky. 
In future work, in order to take into account the spatial
variation of the foreground spectra,
we will attempt to apply {\fastica} on smaller 
patches of the sky where the assumption of uniform spectral behaviour is
more realistic.
Given the success of our method in demonstrating the presence of residual foregrounds surviving
template-based corrections, we can perhaps utilise the information
(ie. the residuals maps) from a near-global approach 
to identify significant regional spectral variations
for such a local analysis. 

\section*{Acknowledgments}

Some of the results in this paper have been derived using the HEALPix
\citep{gorski_etal_2005} package. We acknowledge the use of the Legacy
Archive for Microwave Background Data Analysis (LAMBDA). Support for
LAMBDA is provided by the NASA Office of Space Science. 
We acknowledge the use of Craig Markwardt's fitting package MPFIT\footnote{http://cow.physics.wisc.edu/$\sim$craigm/idl/fitting.html}.

\appendix
\section{}
\label{appendix}

\begin{figure}
\begin{centering}
\includegraphics[angle=90,width=1.\textwidth]{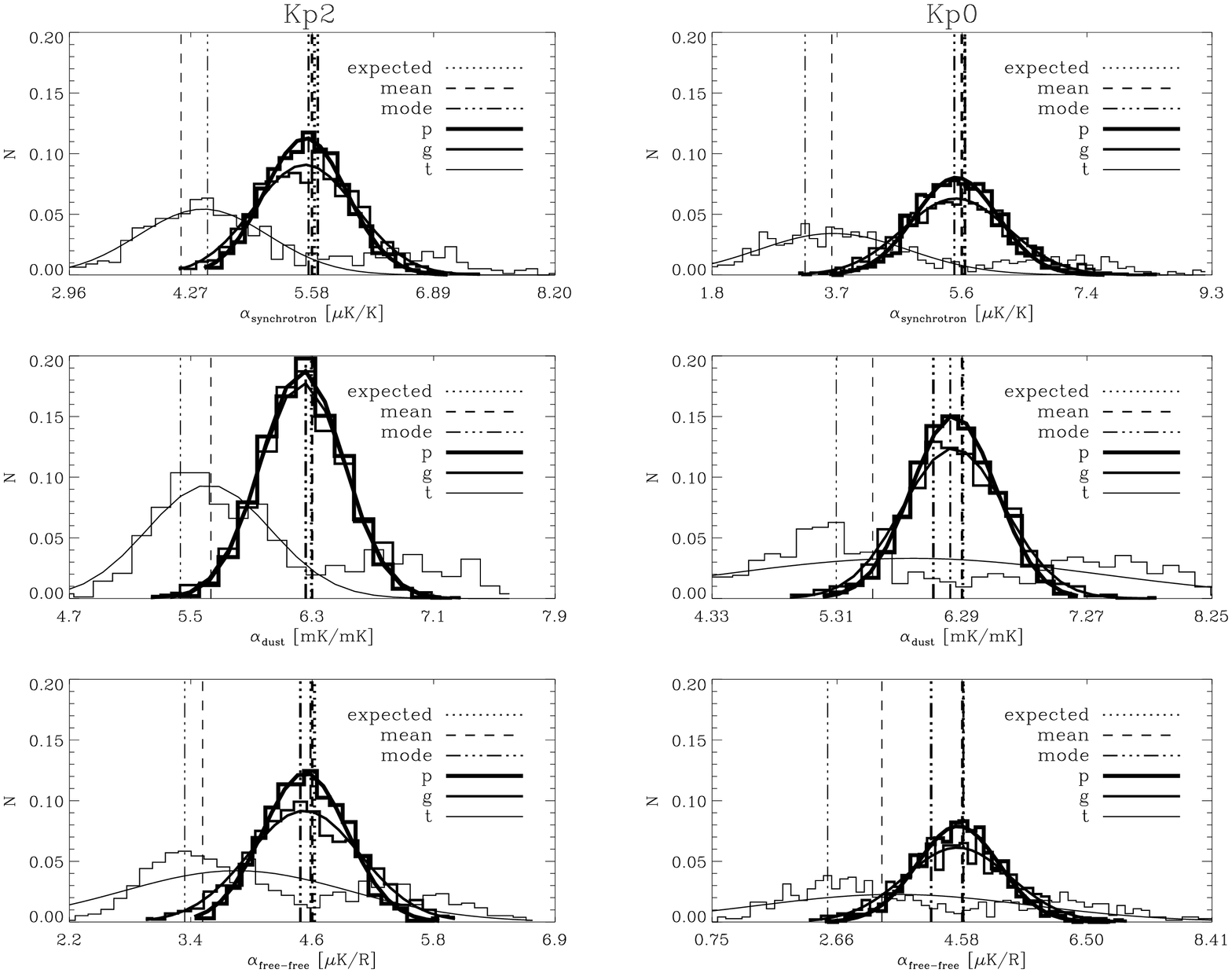}
\caption{\emph{Coupling coefficients distribution determined
from simulations of \WMAP\ K-band data with the p-, g- and t-functions.
The mean (long-dashes and dashed line), the mode (dash-dot and dash-dot-dot line) and
the input values (dotted line) are also shown, together with 
the best-fit Gaussian profiles.}}
\label{figA1}
\end{centering}
\end{figure}

We have performed 1000 simulations of the microwave sky
at each of the 5 \WMAP\ frequencies, each containing a realisation of 
the CMB signal, the Galactic foreground emission, and instrumental
noise appropriate to the specific channel. The study was undertaken
at an effective resolution of $1^{\circ}$.
Each CMB component corresponds to a Gaussian realisation of the 
theoretical best fit CMB angular power spectrum with no running
spectral index as determined by the \WMAP\ first-year 
analysis\footnote{http://lambda.gsfc.nasa.gov/data/map/powspec/wmap\_lcdm\_pl\_model\_yr1\_v1.txt}. 
As verified by \citet{maino_etal_2006} up to $\ell \sim 400$ -- the
scales of interest for 1$^{\circ}$ smoothed data -- the 
differences between the 3 year and 1 year best-fit models are less than 1.5\% and
therefore not expected to bias the results significantly.
The white noise component is simulated for each of the 10 DAs in the
usual manner. That is, for a given pixel a Gaussian random number is selected
with zero mean and a variance corresponding to the 
ratio of the noise variance per observation for that channel 
and the number of observations for the pixel in question.
The CMB component and noise realisations are then combined, and
the maps smoothed. For those frequencies with multiple associated DAs, 
the band averaged skymaps are formed using simple averaging.
For the foreground emission, we added to each simulated frequency map
three Galactic templates as described in section~\ref{data},
appropriately scaled to the frequency in question.
These maps were then used as input data for {\fastica}.

As a convenient point of reference for our \fastica\ results, we
have also included a simple template fitting analysis. In general,
the correlation method can be extended to include various
constraints on the data, eg. fixed dust or free-free spectral
indices. This is the case described by \citet{hinshaw_etal_2007} in
the analysis of the 3 year data of \WMAP\, but we impose no such
constraints here. The best-fit monopole is removed
from each of the sky maps before computing the coefficients, to be
consistent with the  \fastica\  approach.

Using these simulations, we can determine the suitability of the
non-quadratic functions adopted in the {\fastica} analysis for
template fitting. Figure~\ref{figA1} summarises the
situation when using the Haslam 408 MHz sky map as a tracer of the
synchrotron emission. The templates were scaled using the weights from
Table 3 of \citet{bennett_etal_2003}, where for the Q-, V- and W-
bands we have considered the average values.

The distribution of coefficients determined by {\fastica} when
utilising the $t$ function is clearly unsatisfactory -- the
distribution is not Gaussian and has broad asymmetric tails, and is
strongly biassed away from the input values.  In the case of the Kp0
mask, the distribution also shows evidence of bimodality, and for many
simulations the {\fastica} analysis does not even converge.  We
conclude that, when using the $t$ function {\fastica} is unable to
distinguish Galactic emission in the \WMAP\ data as described by
template maps, and we do not include results based on this
non-quadratic function for the analysis of the real data.

In contrast, the behaviour of the $p$ and $g$ functions
for determining the coupling coefficients seems to be suitable
for our purposes.
Although we show results only for the K-band simulations,
we find that at all frequencies the scaling factors are
very well described by Gaussian distributions --
in fact the corresponding measures of
the skewness and kurtosis values are small in all cases.
The mean values for the simulated distributions
of coefficients are very similar to the inputs,
and there is no evidence of significant bias.
This is true also for the mode if we consider $p$, but
with $g$ the mode is slightly smaller than the input value for the dust and
free-free emission: indeed, it could explain partially the difference
between the coefficients obtained in the real analysis.
However, the main difference between the two functions is that
the distribution determined with $g$ is broader than with 
the $p$ function: corresponding to a larger statistical
uncertainty for analyses made with the former.

It is likely that some of the performance features of the different
non-linear functions are related to the statistical features
of the physical emission as traced by the template sky maps.
Not surprisingly, the results indicate that the template fits
using {\fastica} are sensitive to the extent of the mask applied to
the data. For a narrower Galactic exclusion region,
the template fits have smaller errors. This is certainly in part
due to simple \lq sample variance' type arguments, but
may also be connected with the degree to which bright foreground
emission near to the applied cut can stabilise the fits.
Variations in the fitted amplitudes of the foreground components
against the observed data as a function of mask may also reveal genuine
changes in the physical properties and spectral behaviour of the
foregrounds with latitude.

\begin{figure}
\begin{centering}
\includegraphics[angle=90,width=1.\textwidth]{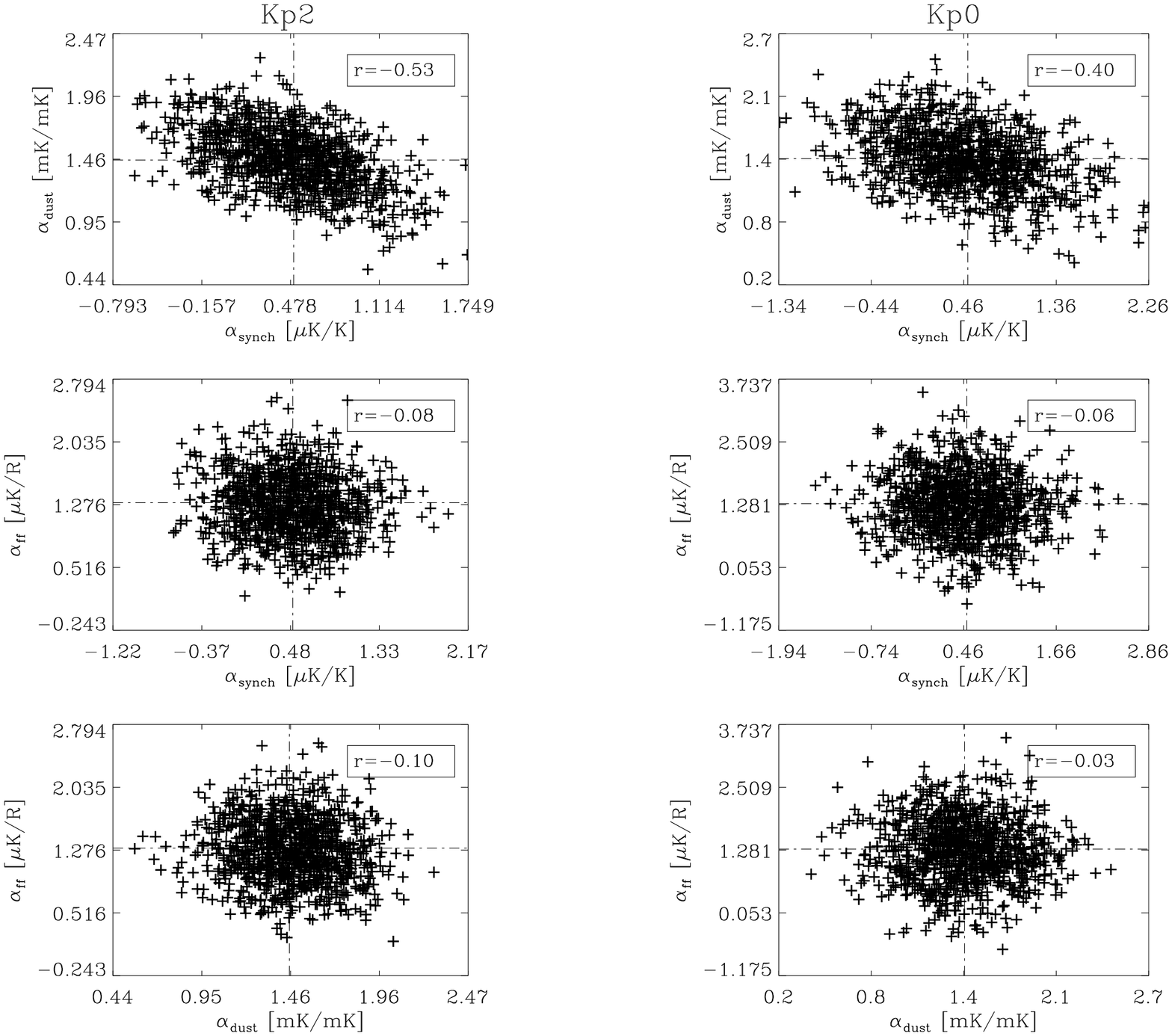}
\caption{\emph{Scatter plots of the coupling coefficients obtained
by {\fastica} with the $p$-function using the simulations performed using the Haslam map as synchrotron
template.
We show as an example the results in the Q-band.
There is very little change in the correlation properties as
a function of frequency.}}
\label{figA2}
\end{centering}
\end{figure}

In Figure~\ref{figA2}
we have plotted the fitted Q-band amplitudes for one template against another
for the various template permutations in order to study the
significance of cross-talk between the maps.
There is an apparent anti-correlation between the synchrotron and the dust
coefficients (as specified by the linear correlation coefficient $r$), 
but no clear evidence for such behaviour between
synchrotron and free-free or free-free and dust. 
This behaviour is essentially identical for all of the \WMAP\ frequencies.
Some care should then be exercised in interpreting the coefficients for
the former two components. 

\begin{figure}
\begin{centering}
\includegraphics[angle=90,width=1.\textwidth]{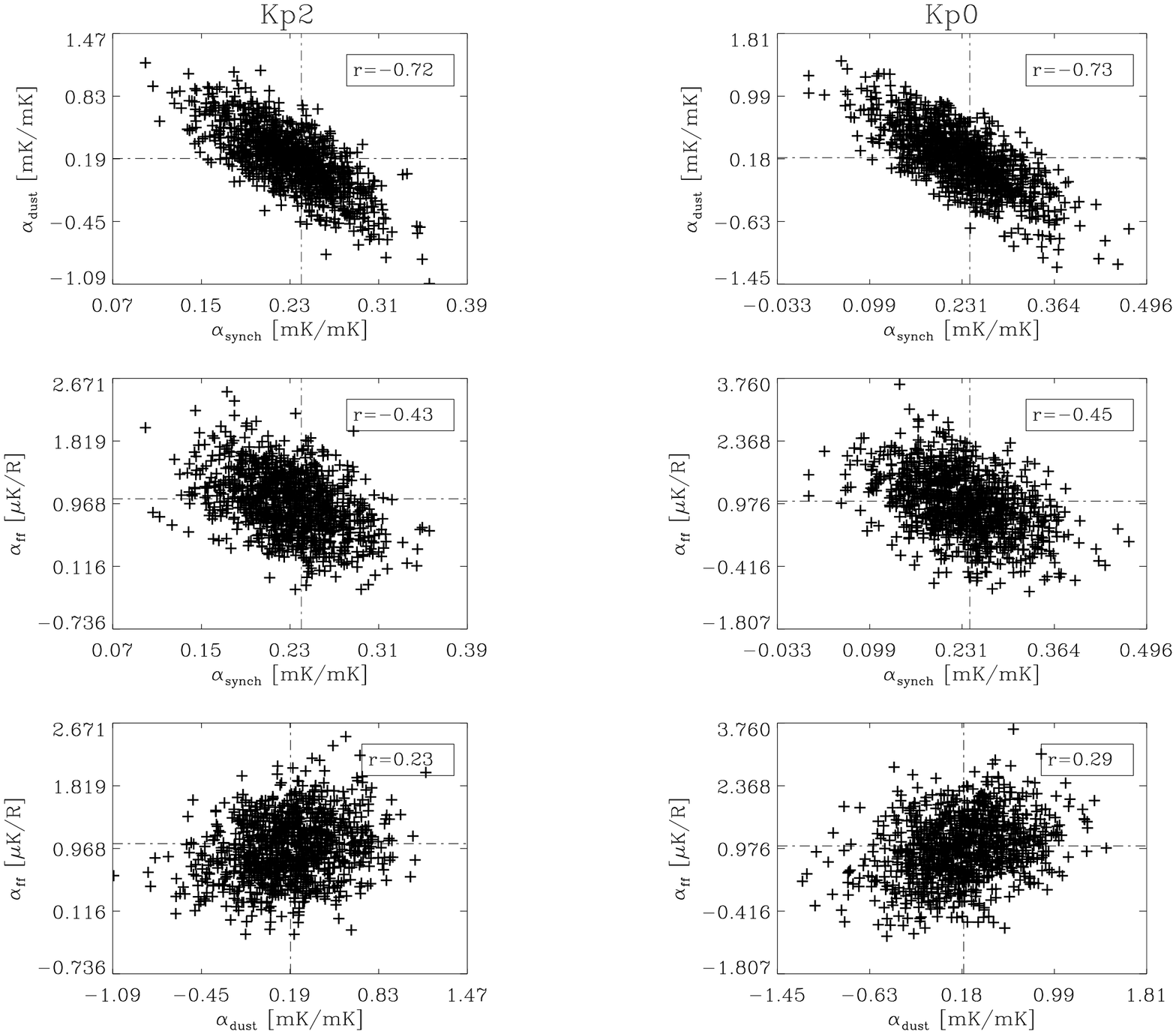}
\caption{\emph{Scatter plots of the coupling coefficients obtained
by {\fastica} with the $p$-function using the simulations performed with the K-Ka map as synchrotron
template.
We show as an example the results in the Q-band.
There is very little change in the correlation properties as
a function of frequency.}}
\label{figA3}
\end{centering}
\end{figure}

We have repeated the simulations and analysis using the \WMAP\
K-Ka map as the synchrotron template, in this case only for the Q-,
V- and W-bands. The templates were scaled using the coefficients from
\citet{hinshaw_etal_2007}, but adopting the mean of the coefficients
over the differencing assemblies.
The distributions of the coefficients are again symmetric and well
fitted by Gaussians as in the case of the Haslam template.
Furthermore, the results are consistent with the previous ones
concerning the dependence of the coefficients on the {\fastica}
functions.  
The analogous scatter plots between pairs of coefficients are shown in
Figure~\ref{figA3}.
This time,the synchrotron coefficients are anti-correlated with
the free-free results. Indeed, this
should be expected since the template must contain both synchrotron
and free-free emission. However, the synchrotron coefficients are 
more strongly anti-correlated with the dust
values, which is easily interpreted as being due to the 
presence of anomalous dust correlated emission in the K-Ka template.
This factor is not taken into account by \citet{hinshaw_etal_2007}
in their constrained template analysis.

\begin{figure}
\begin{centering}
\includegraphics[angle=90,width=1.\textwidth]{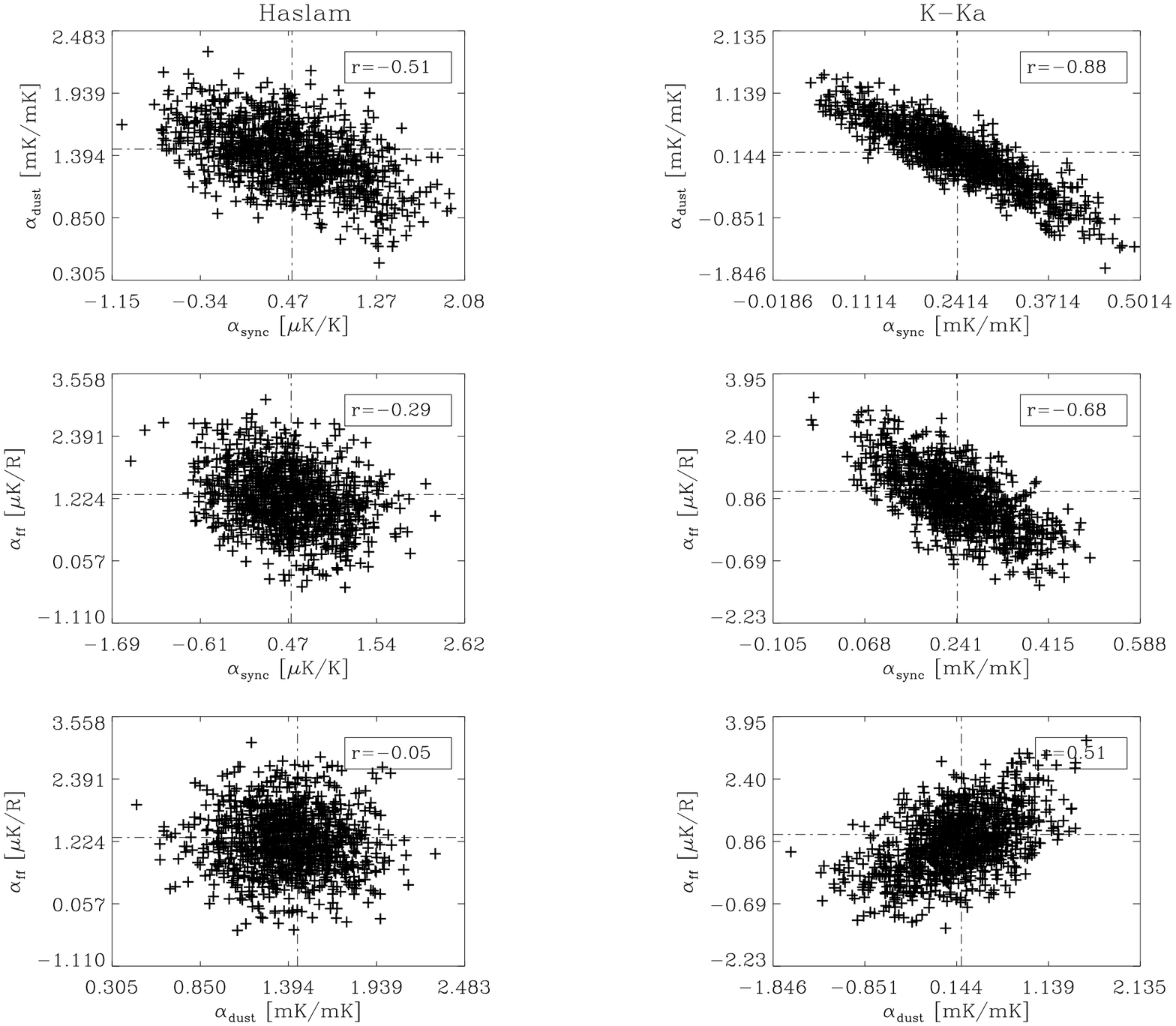}
\caption{\emph{Scatter plots of the scaling factors obtained
via the simple $\chi^2$ method on $Kp2$ sky coverage with simulations performed using
either the Haslam map (left column) or the K-Ka map (right column) as the synchrotron
template. 
We show as an example the results in the Q-band.
There is very little change in the correlation properties as
a function of frequency.}}
\label{figA4}
\end{centering}
\end{figure}

Finally, we performed the same statistical study with the simple $\chi^2$
minimisation in order to compare the performances of the two methods.
The results depend on the template considered.  For the synchrotron
and free-free emission, the uncertainties are larger than the
{\fastica} values obtained with either the $p$- or $g$-function,
substantially so in the former case.  For the dust emission they are
essentially equivalent.  It is therefore clear that {\fastica} is a
reliable method to perform template fitting, in some cases
outperforming the conventional $\chi^2$ minimisation.  This is
particularly true for $p$, and we could conclude that it is the most
stable non-linear function, although this is not a definitive
argument. Figure~\ref{figA4} shows scatter plots
between pairs of recovered Q-band coefficients when using either the Haslam or K-Ka
maps as synchrotron templates. It is intersting to note that there is
a hint of anti-correlation between the synchrotron and free-free
coefficients using the Haslam template, but that this becomes
significant when the K-Ka template is employed. Moreover, a
significant correlation between free-free and dust is seen in the
latter case. The simple $\chi^2$ method appears to demonstrate
more cross-talk between the coefficients of the different components
than seen for the {\fastica} analysis. This may help to understand
some of the differences seen in the results obtained with the observed
data.

\label{lastpage}
\end{document}